\documentclass[12pt]{article}

\usepackage[utf8]{inputenc}
\usepackage[round]{natbib}
\usepackage{mathrsfs}   
\usepackage{amssymb,amsthm,amsmath,amsfonts,eurosym, geometry,ulem,graphicx,caption,color,setspace,comment,footmisc,caption,pdflscape,subfigure,array,hyperref,xcolor,bm, dcolumn, float, bbm, eqnarray}

\usepackage{array}
\newcolumntype{H}{>{\setbox0=\hbox\bgroup}c<{\egroup}@{}} 

\usepackage[affil-it]{authblk}
\usepackage{threeparttable}                     
\usepackage{multirow}             
\usepackage[toc,page]{appendix}
\usepackage{mathtools}
\usepackage[round]{natbib}
\usepackage{mathrsfs}   

\newlength{\dhatheight}

\newcommand\T{\rule{0pt}{2.6ex}}
\newcommand\B{\rule[-1.2ex]{0pt}{0pt}}

  \usepackage{tikz} 
  \usepackage{pgfplots}
  \usetikzlibrary{pgfplots.dateplot}
  \usepackage{pgfplotstable}
  \pgfplotsset{compat=1.17}
  \usetikzlibrary{intersections,patterns,pgfplots.fillbetween}
\usetikzlibrary{decorations.pathreplacing} 

 \usepackage{algorithm} 

\usepackage{amsthm}
\newcommand{\E}{\operatorname{\mathbb{E}}} 
\newcommand{\R}{\operatorname{\mathbb{R}}} 
\newcommand{\p}{\operatorname{\mathbb{P}}} 
\newcommand{\N}{\operatorname{\mathbb{N}}} 
\newcommand{\C}{\operatorname{\pazocal{C}}} 

\DeclareMathOperator*{\argmin}{arg\,min}
\DeclareMathOperator*{\supp}{supp}

\DeclareMathOperator*{\sign}{sign}

\DeclareMathOperator*{\var}{Var}
\DeclareMathOperator*{\cov}{Cov}
\DeclareMathAlphabet{\pazocal}{OMS}{zplm}{m}{n}
\DeclareMathOperator*{\lip}{Lip}
\DeclareMathOperator*{\as}{a.s.}

\newtheorem{theorem}{Theorem}

\newtheorem{assump}{Assumption}
\newtheorem{defin}{Definition}
\newtheorem{lem}{Lemma}

\newcolumntype{L}[1]{>{\raggedright\let\newline\\arraybackslash\hspace{0pt}}m{#1}}
\newcolumntype{C}[1]{>{\centering\let\newline\\arraybackslash\hspace{0pt}}m{#1}}
\newcolumntype{R}[1]{>{\raggedleft\let\newline\\arraybackslash\hspace{0pt}}m{#1}}

\begin{document}

\title{Moran's $I$ 2-Stage Lasso: for Models with Spatial Correlation and Endogenous Variables.\thanks{We are thankful to Abhimanyu Gupta and Hans-Martin Krolzig for discussions and suggestions.}}

\author{By Sylvain Barde, Rowan Cherodian and Guy Tchuente\thanks{Barde: University of Kent; Cherodian (corresponding author): University of Sheffield, email: \texttt{r.cherodian@sheffield.ac.uk}; Tchuente: Purdue University}}
\date{\today}

\maketitle

\begin{abstract}
We propose a novel estimation procedure for models with endogenous variables in the presence of spatial correlation based on Eigenvector Spatial Filtering. The procedure, called Moran's $I$ 2-Stage Lasso (Mi-2SL), uses a two-stage Lasso estimator where the Standardised Moran's $I$ is used to set the Lasso tuning parameter. Unlike existing spatial econometric methods, this has the key benefit of not requiring the researcher to explicitly model the spatial correlation process, which is of interest in cases where they are only interested in removing the resulting bias when estimating the direct effect of covariates. We show the conditions necessary for consistent and asymptotically normal parameter estimation assuming the support (relevant) set of eigenvectors is known. Our Monte Carlo simulation results also show that Mi-2SL performs well against common alternatives in presence of spatial correlation. Our empirical application replicates \citet{ck16} instrumental variables estimates using Mi-2SL and shows that in that case Mi-2SL can boost the performance of the first stage.
\end{abstract}

\newpage

\section{Introduction}



The main aim of structural economic modeling is to explain the evolution of endogenous variables of interest, given fundamental processes such as productivity, taste, and policy. It has long been known that Ordinary Least Squares (OLS) estimation of the coefficients of such endogenous variables is invalidated by endogeneity bias and that instrumental variables (IV) offer a way around this problem \citep{wright28}. This paper considers the case where the researcher is similarly interested in estimating the parameters on endogenous variables, but where in addition both the structural equation being estimated and the endogenous variables themselves spatial processes based on a given spatial weights matrix (SWM).\footnote{A spatial weights matrix is an $n\times n$ matrix that describes the pair-wise relationship between each of the $n$ cross-sectional units.} Crucially, while SWM is assumed to be known, we do assume that the exact functional forms of the spatial processes are \textit{unknown}, and possibly include higher-order powers of the SWM. Because the researcher is only interested in estimating the direct effect of the right-hand-side variable(s), the corresponding spatial parameters are thus considered nuisance parameters.

This setup is arguably a realistic situation in applied research: testing for cross-sectional/ spatial dependence is relatively easy, for example using a Moran's $I$ test \citep{moran50}, but determining the exact form of the spatial process is much more challenging, and might not form the focus of the research. Similarly, spatial dependence and endogeneity are common in many economic models. Some examples include modelling the relationship between economic growth and energy consumption or pollution, employment and migration, and the effect of policing on crime. Many papers in the econometrics literature have shown how to incorporate endogenous variables into a given spatial model.\footnote{Some recent examples include \citet{H18_JBES, J16_ET, XL13_ER, FLG08}.} The Generalised Method of Moment (GMM) based estimation techniques such as Generalised Spatial Two-Stage Least Squares (GS2SLS) are commonly used by applied researchers when estimating a spatial model with an endogenous variable. However, to use any of the proposed GMM-based estimation techniques, the researcher must specify (1) a spatial economic model and (2) define the spatial structure, i.e., the SWM. A misspecified model will yield inconsistent estimates, and this problem is more acute if the SWM is also misspecified \citep{lsK_sens14}.

Given that the spatial process is assumed to be of a lesser interest to the researcher than the direct economic impact of the endogenous variable, i.e. the spatial parameters are considered nuisance parameters, we propose relying on the Eigenvector Spatial Filtering (ESF) approach developed by \cite{grif2000, grif03}. This has the key advantage of being agnostic to the underlying functional form of the spatial process. Instead of explicitly modelling the underlying spatial process, ESF uses a subset of eigenvectors from the SWM as controls in a linear regression framework to control of the spatial dependence, removing the need to specify and estimate a spatial process.

Leaving aside the issue of endogeneity for a moment, the main downside of ESF is that estimation using the full set of eigenvectors is infeasible using OLS. Given $k$ covariates, the addition of the $n$ eigenvectors produced by the spectral decomposition of the SWM necessarily produces a rank-deficient Gram matrix with $n+k$ parameters and $n$ observations. This problem can be mitigated by making a sparsity assumption, i.e. assuming that only a subset of the eigenvectors are relevant and will have non-zero coefficients. This generates a separate problem, however, which is the selection of the relevant subset of eigenvectors. To solve this selection problem, we propose using a Lasso-based procedure that uses information contained in the Moran's $I$ statistic to determine a point estimate for the Lasso tuning parameters. The proposed estimator, called Moran's $I$ two-stage Lasso (Mi-2SL), is a three-step procedure: the first and second stages of a general two-stage least squares (2SLS) specification are separately estimated by using this Moran's $I$ based Lasso, in order to extract the relevant eigenvectors. The union of the two sets of selected eigenvectors is then used to provide supplementary covariates in a standard 2SLS regression. This 2SLS specification deals with the endogenous variables, with the additional eigenvectors selected via Moran's $I$ based Lasso dealing with the (weak) cross-sectional dependence.\footnote{We will use the terms cross-sections dependence and spatial dependence interchangeably.}

Several studies have already used two-stage Lasso procedures in a spatial setting. For example, \citet{peng19} estimates a spatial autoregressive model (SAR) by a two-stage Lasso procedure to allow heterogeneous peer effects and the identification of the influential individuals in a network. As both stages are high-dimensional, they are both estimated by Lasso. \citet{Ahren15} estimate the effect of conflict risk on economic growth using \citet{bcch12} two-stage procedure, where Lasso estimates the high-dimensional first stage and the second is a low-dimensional panel SAR model. Additionally, \citet{arhbha15, LS16_SWM, LS20_SWM} all use two-stage Lasso-based procedures to estimate/select a SWM. We are the first, however, to consider a two-stage Lasso procedure for ESF.


The specific contribution we bring is to derive theoretical results on consistent and asymptotically normal parameter estimation. Proving consistency and asymptotically normality is tricky: as the eigenvectors are derived from the SWM, which itself encodes the pair-wise dependence between the observations, one cannot rely on the standard assumption of row-wise independence. To get around this problem we rely instead on the \citet{LT_KMS20} notion of $\psi$-dependence and corresponding limit theorems to derive our results. These theoretical results are supported by a set of Monte Carlo simulations, where the estimator is tested against competing methodologies for varying degrees of correlation between the first and second-stage errors as well as varying levels of spatial dependency in the covariates. The analysis shows that Mi-2SL performs well relative to competitors in small samples, and out-performs them in terms of bias and mean squared errors in the presence of spatially correlated covariates.

Finally, as a motivating application, we apply our methodology to \citet{ck16}, who analyse the impact of Mexican worker mobility on local labour market outcomes of natives in the US, using a standard IV strategy to correct for endogeneity. Despite having an explicit spatial dimension in their data, their analysis does not allow for spatial dependence in their specification. A standardised Moran's $I$ test on the first and second-stage residuals indicates significant spatial correlation for most demographic groups, with a higher spatial correlation level in the first stage than the second. This forms an idea use-case for Mi-2SL, as the functional form of the spatial process is uncertain, and it is not the main focus of the research question. We re-estimate their model using Mi-2SL to account for the unknown spatial structure and find that while Mi-2SL does not change the overall conclusion of \citet{ck16}, it substantially improves the strength of the Bartik instrument in the first stage, thus improving the precision of the second stage estimates.

The rest of the paper is structured as follows. Section \ref{sec2} presents the underlying structural model, the notation, and the proposed the Mi-2SL procedure. In section \ref{theory2} we drive the theoretical properties of Mi-2SLS under perfect selection. Section \ref{sims2} provides Monte Carlo studies to evaluate the finite sample properties of the proposed estimator and in Section \ref{app_ck16}, we apply the proposed procedure to \citet{ck16}. Finally, Section \ref{conc} offers our concluding remarks.

\section{Structural model and estimation procedure}\label{sec2}
\subsection{Underlying structural model}\label{model}

Consider the following structural equation where the endogenous $n\times 1$ vector $\bm{y}$ which depends on an $n\times k_1$ matrix of exogenous regressors $\bm{X}_1$, an $n\times 1$ endogenous vector $\bm{x}_2$ and follows some spatial process:
\begin{equation}
  \label{stage2}
      \bm{y} =\bm{X}_1\bm{\beta}_{1,0}+\bm{x}_2\beta_{2,0}+f(\bm{W},\bm{y},\bm{X}_1)+\bm{\varepsilon}
\end{equation}
where $f(\bm{W},\bm{y},\bm{X}_1)$ is a linear combination of spatial lags of $\bm{y}$ and $\bm{X}_1$ obtained with $\bm{W}$, a $n \times n$ symmetric weights matrix, and $\bm{\varepsilon}$ is an $n\times 1$ vector of innovations. $f(\bm{W},\bm{y},\bm{X}_1)$ is allowed to contain higher-order spatial lags $\bm{W}^i\bm{y}$ and $\bm{W}^i\bm{X}_1$ with $i>1$. An example of a common special case of this process is:
\begin{equation}
    \bm{y} =  \bm{X}_1\bm{\beta}_{1,0} + \bm{x}_2\beta_{2,0} +\sum^p_{i=1}\bm{W}^i\bm{y}\rho_{i,0}+ \bm{W}\bm{X}_1\bm{\psi}_0+\bm{\varepsilon},   \label{ydgp}
\end{equation}
where $\rho_{i,0}$'s and $\bm{\psi}_0$ are unknown parameters that represent the degree of spatial correlation in the endogenous variable $\bm{y}$ and the predetermined exogenous variables $\bm{X}_1$ with moment conditions $\E[\bm{X}_1'\bm{\varepsilon}]=0$ and $\E[(\bm{W}\bm{X}_1,\bm{X}_1)'\bm{\varepsilon}]=0$. The exact spatial process is unknown, in the sense that some of these spatial parameters, including $p$, are allowed to be zero-valued.\footnote{The data generating process of $\bm{y}$ could also include spatial autoregressive disturbances; however this is excluded from the model for simplicity.}

The regressor $\bm{x}_2$ in \ref{stage2} is endogenous, in the sense that $\E(\bm{x}_{2}'\bm{\varepsilon} )\neq 0$, and $\beta_{2,0}$ is the parameter of interest to the researcher. The extension to the case where $\bm{x}_2$ is a matrix is straightforward and omitted for simplicity. We assume that $\bm{x}_2$ also follows some unknown spatial process:
\begin{equation}
    \bm{x}_2 = \bm{X}_1\bm{\zeta}_{1,0}+\bm{Z}_2\bm{\zeta}_{2,0} + g(\bm{W},\bm{x}_2, \bm{X}_1,\bm{Z}_2) + \bm{u}_2
\end{equation}
where $\bm{Z}_2$ is a $n\times q$ matrix of instrument variables with $q\geq 1$ and moment conditions $\E(\bm{Z}_2'\bm{\varepsilon})= 0$. $\bm{u}_2$ is a vector of disturbances with $\E[(\bm{X}_1,\bm{Z}_2,\bm{W}\bm{X}_1,\bm{W}\bm{Z}_2)'\bm{u}_2]=0$ and $\E[(\bm{X}_1,\bm{Z}_2,\bm{W}\bm{X}_1,\bm{W}\bm{Z}_2)'\bm{\varepsilon}]=0$. Again, the spatial process $g(\bm{W},\bm{x}_2, \bm{X}_1,\bm{Z}_2)$ is some linear combination of spatial lags of $\bm{x}_2$, $\bm{X}_1$ and $\bm{Z}_2$, obtained with $\bm{W}$. An example of such a process is:
\begin{equation}
   \bm{x}_2 = \bm{X}_1\bm{\zeta}_{1,0}+\bm{Z}_2\bm{\zeta}_{2,0} +  \bm{W}\bm{X}_1\bm{\zeta}_{3,0}+\bm{W}\bm{Z}_2\bm{\zeta}_{4,0} + \sum^l_{i=1}\bm{W}^i\bm{x}_2\zeta_{i,5,0} + \bm{u}_2  \label{xdgp}
\end{equation}

Let $N_n=N=\{1, \ldots, n\}$ be the set of cross-sectional unit indices with $n\in\N$ denoting the number of observations. For reasons of generality, we allow the elements of
$\bm{\varepsilon}=\bm{\varepsilon}_n$, $\bm{y}=\bm{y}_n$, $\bm{W}=\bm{W}_{n}$, $\bm{Z}_2=\bm{Z}_{2,n}$, $\bm{u}_2=\bm{u}_{2,n}$,
 $\bm{X}_1=\bm{X}_{1,n}$ and $\bm{x}_2=\bm{x}_{2,n}$ to be dependent on $n$, that is to form triangular arrays. However, to simplify the notation, the $n$ index is omitted.

Equation \eqref{ydgp} contains two sources of endogeneity, first
$\bm{x}_2$ because $\E(\bm{u}_2'\bm{\varepsilon} )\neq 0$, which implies $\E(\bm{x}_2'\bm{\varepsilon} )\neq 0$. Second, $\bm{y}$ itself is endogenous as it appears on both sides of \eqref{ydgp}, via $\bm{W}^i\bm{y} \; \forall i$. Both sources of endogeneity cause the OLS estimate of $\bm{\beta}_0=(\bm{\beta}_{1,0}, \beta_{2,0})'$
to be inconsistent ($\hat{\bm{\beta}}_{ols}\not\to_p \bm{\beta}_0$).

Substituting \eqref{xdgp} into \eqref{ydgp} gives the reduced forms for $\bm{y}$:
\begin{align}
   \bm{y}=&\bm{S}_1^{-1}\left(\bm{X}_1\bm{\beta}_{1,0}+\bm{S}_2^{-1}(\bm{X}_1\bm{\pi}_{1,0} + \bm{Z}_2\bm{\pi}_{2,0}+ \bm{W}\bm{X}_1\bm{\pi}_{3,0}+\bm{W}\bm{Z}_2\bm{\pi}_{4,0}) +\bm{WX}_1\bm{\psi}_0+\bm{d}\right) \label{rfy}
\end{align}
where $\bm{\pi}_{1,0}=\beta_{2,0}\bm{\zeta}_{1,0}$, $\bm{\pi}_{2,0}=\beta_{2,0}\bm{\zeta}_{2,0}$, $\bm{\pi}_{3,0}=\beta_{2,0}\bm{\zeta}_{3,0}$, $\bm{\pi}_{4,0}=\beta_{2,0}\bm{\zeta}_{4,0}$, $\bm{d}=\bm{S}_2^{-1}\bm{u}_2\beta_{2,0} + \bm{\varepsilon}$ and both $\bm{S}_1\equiv(\bm{I}-\sum^p_{i=1}\rho_{i,0}\bm{W}^i)$, and $\bm{S}_2\equiv(\bm{I}-\sum^l_{i=1}\bm{W}^i\zeta_{i,3,0})$ are non-singular.

\subsection{Moran's $I$ 2-Stage Lasso}\label{miv}

The existence of valid instruments $\bm{Z}_2$ for the endogenous $\bm{x}_2$ implies that we can deal with the problem of endogeneity, leaving the key challenge of controlling for the unknown underlying spatial processes in \eqref{ydgp} and \eqref{xdgp}. Even if the exact underlying spatial process were known, estimation of \eqref{ydgp} would be feasible, albeit non-trivial. One method would be to first estimate \eqref{xdgp} by GS2SLS, first developed by \citet{kelpru98} and extended by \citet{DEP19_HOG2SGS} to allow for higher-order spatial lags, which would use higher order spatial lags of the exogenous variables in \eqref{xdgp} as instruments for $\bm{W}^i\bm{x}_2 \; \forall i$. The resulting fitted values can then be used to estimate \eqref{ydgp}. GS2SLS has the advantage that it can be easily extended to include other right-hand-side endogenous variables. However, the procedure requires that the researcher specify which spatial parameters to estimate, and given this extra layer of estimation, the standard GS2SLS standard errors would be invalid.

Given the additional assumed uncertainty regarding the true functional form of the spatial process in the model, we propose using eigenvectors $\bm{E}_{n}=\bm{E}$ from a spectral decomposition of $\bm{W}$ to represent $f(\bm{W},\bm{y},\bm{X}_1)$ and $g(\bm{W},\bm{x}_2, \bm{X}_1,\bm{Z}_2)$ i.e. $f(\bm{W},\bm{y},\bm{X}_1)=\bm{E}\bm{\gamma}_{y,0}$ and $g(\bm{W},\bm{x}_2, \bm{X}_1,\bm{Z}_2)=\bm{E}\bm{\gamma}_{x,0}$ where $\bm{\gamma}_{y,0}$ and $\bm{\gamma}_{x,0}$ are vectors of unknown constants. This methodology has the key advantage that it is agnostic to the exact form of $f(\bm{W},\bm{y},\bm{X}_1)$ and $g(\bm{W},\bm{x}_2, \bm{X}_1,\bm{Z}_2)$, including the presence of higher-order lags, stemming from the spectral property that the eigenvectors from $\bm{W}$ and $\bm{W}^i$ $\forall i \in \mathbb{Z}^+$ are the same. Using this linear representation, one could in principle estimate the following system instead of \eqref{ydgp} and \eqref{xdgp}:
\begin{align}
 \bm{y} & = \bm{G\Upsilon}_0 +\bm{\varepsilon} \label{esf} \\
 \bm{x}_2 & = \bm{Z\zeta}_{0} + \bm{u}_2  \label{xdgpE}
\end{align}
where $\bm{G}=[\bm{X}_1,\bm{x}_2,\bm{E}]$, $\bm{\Upsilon}_0=[\bm{\beta}_{1,0},\beta_{2,0}, \bm{\gamma}_{y,0}]'$, $\bm{Z}=[\bm{X}_1,\bm{Z}_2, \bm{E}]$
and $\bm{\zeta}_0=[\bm{\zeta}_{1,0}, \bm{\zeta}_{2,0}, \bm{\gamma}_{x,0}]'$ with $\E[\bm{G}'\bm{\varepsilon}]=0$ and $\E[\bm{Z}'\bm{u}_2]=0$.

The practical obstacle is that \eqref{esf} and \eqref{xdgpE} are both high-dimensional linear regressions, as in each equation the number of parameters is greater than the number of observations. This means both the (re-scaled) Gram matrices $\bm{G}'\bm{G}/n$ and $\bm{Z}'\bm{Z}/n$ are necessarily rank deficient. Thus, neither \eqref{esf} nor \eqref{xdgpE} cannot be estimated by OLS nor \eqref{esf} by 2SLS. \cite{grif2000} argues, however, that in most cases only a subset of eigenvectors are relevant to the data generating process (DGP) of $\bm{y}$ and $\bm{x}_2$, i.e. the parameter vectors $\bm{\gamma}_{y,0}$ and $\bm{\gamma}_{x,0}$ are sparse. The intuition behind this sparsity assumption is each of the $n$ eigenvectors can be viewed as an orthogonal spatial pattern, and only a specific subset of these patterns are relevant to the DGP of $\bm{y}$ and $\bm{x}_2$ \citep{grif03}. Thus, the estimation problem turns into a selection problem.

We propose addressing this selection problem with an extension of the Moran's $I$ based Lasso first proposed in \citet{c23}. This procedure considers a single structural equation where all the covariates are exogenous, i.e., \eqref{ydgp} with $\beta_2=0$. It only penalises the $\bm{\gamma}_y$ coefficients on the eigenvectors $\bm{E}$ and set the Lasso tuning parameter to $z^{-2} \; \forall \; z\neq 0$ where $z$ is the standardised Moran's $I$ ($z$) of the residual $\hat{\bm{h}}=\bm{M_Xy}$, with $\bm{M_X}=\bm{I}-\bm{X}_1(\bm{X}_1'\bm{X}_1)^{-1}\bm{X}_1'$.

\begin{equation}\label{zmi}
    z	= \Bigg( \frac{m-\E[m]}{\sqrt{\var(m)}} \Bigg)
\end{equation}
with
\begin{align*}
  m &= \frac{\hat{\bm{h}}'\bm{W}\hat{\bm{h}}}{\hat{\bm{h}}'\hat{\bm{h}}}, \\
  \E[m] &= \frac{tr(\bm{M_XWM_{X}})}{n-k}, \\
 \var(m) &= \frac{2\bigg((n-k)tr\big((\bm{M_{X}WM_{X}})^2\big)- \big[tr(\bm{M_{X}WM_{X}})\big]^2\bigg)}{(n-k)^2(n-k-2)}
\end{align*}

Given that the aim of ESF is to directly control for spatial correlation patterns in the regression, the intuition behind calibrating the tuning parameter this way is that when the level of spatial correlation in the residuals is low, only a small set of eigenvectors is necessary, thus a high level of regularization (large tuning parameter) is required. In contrast, when the level of spatial correlation is high, a larger set of eigenvectors will be necessary, thus a low level of regularization (small tuning parameter) is required. As $z$ gives a large value when the overall correlation is high and small values when the overall correlation is low, they propose using the inverse square of the standardised Moran's $I$ as the tuning parameter.\footnote{A positive tuning parameter is required for the Lasso solution to be unique. Thus, the squared value of $z$ is used.}

Our proposed Moran's $I$ 2-stage Lasso (Mi-2SL) procedure, outlined in Algorithm \ref{Mi-2SLalg}, can handle both endogenous covariates and cross-sectional dependence. The procedure is straightforward: first a spectral decomposition of the SWM is performed to get the candidate set of eigenvectors. The standardised Moran's $I$ on the naïve first stage residuals (ignoring the spatial correlation) provides the tuning parameter for a Lasso (or post-Lasso) estimation of \eqref{Mi-2SL1st} to get $\hat{\bm{x}}_2$ as well as the selected eigenvectors $\hat{\bm{E}}_x$. Subsequently, $\hat{\bm{x}}_2$ is used instead of $\bm{x}_2$ to calculate standardised Moran's $I$ for the naïve second stage residuals (ignoring the spatial correlation), which serves as the tuning parameter for a lasso estimation of \eqref{Mi-2SL2nd}, providing a second set of selected eigenvectors $\hat{\bm{E}}_y$. As a final step, $\beta_2$ is estimated by standard 2SLS using the union of $\hat{\bm{E}}_x$ and $\hat{\bm{E}}_y$ as additional controls.

\begin{algorithm}[t]
\caption{Mi-2SL Algorithm pseudocode}
\label{Mi-2SLalg}
\begin{enumerate}
  \item Decompose the SWM to get the candidate set of eigenvectors $\bm{E}$.
  \item Estimate naïve first stage residuals $\hat{\bm{r}}= \bm{M}_{H} \bm{x}_2$ where $\bm{M}_H = \bm{I}-\bm{H}(\bm{H}'\bm{H})^{-1}\bm{H}'$ and $\bm{H}=(\bm{X}_1,\bm{Z}_2)$
  and calculate the standardised Moran's $I$ of $\hat{\bm{r}}$, denoted $z_x$.
    \item Estimate:
   \begin{equation}\label{Mi-2SL1st}
     [\hat{\bm{\zeta}}_{1},\hat{\bm{\zeta}}_{2},\hat{\bm{\gamma}}_{x}]\in \argmin \{ ||\bm{x}_2-\bm{X}_1\bm{\zeta}_1-\bm{Z}_2\bm{\zeta}_2-\bm{E}\bm{\gamma}_{x}||^2_2 +z_{x}^{-2} ||\bm{\gamma}_{x}||_1 \}
   \end{equation}
   Use the Lasso or post-Lasso estimates of \eqref{Mi-2SL1st}, save the fitted $\hat{\bm{x}}_2$ and selected set of eigenvectors $\hat{\bm{E}}_x$.
     \item Estimate naïve second stage residuals $\hat{\bm{h}}= \bm{M}_{\hat{\bm{X}}} \bm{y}$ where $\bm{M}_{\hat{\bm{X}}} =\bm{I}-\hat{\bm{X}}(\hat{\bm{X}}'\hat{\bm{X}})^{-1}\hat{\bm{X}}'$
      and $\hat{\bm{X}}=(\bm{X}_1,\hat{\bm{x}}_2)$ and calculate the standardised Moran's $I$ of $\hat{\bm{h}}$, denoted $z_y$.
   \item Estimate:
  \begin{equation}\label{Mi-2SL2nd}
    [\hat{\bm{\beta}}_{1},\hat{\beta}_{2},\hat{\bm{\gamma}}_{y}]\in \argmin \{ ||\bm{y}-\bm{X}_1\bm{\beta}_1-\hat{\bm{x}}_2\beta_2-\bm{E}\bm{\gamma}_{y}||^2_2 +z_{y}^{-2} ||\bm{\gamma}_{y}||_1 \}
  \end{equation}
  and save the selected set of eigenvectors $\hat{\bm{E}}_y$.
  \item Estimate $\beta_2$ by 2SLS using $\hat{\bm{E}}_x \cup \hat{\bm{E}}_y$ as additional controls.
\end{enumerate}
\end{algorithm}

\section{Theoretical results}\label{theory2}

\subsection{Assumptions}

We will now derive some theoretical properties of the proposed Mi-2SL procedure. This requires two sets of assumptions, the first of which applies to the underlying data generating processes \eqref{ydgp} and \eqref{xdgp}.

\begin{assump}[Regularity of DGP]\label{as:spatial2}
$\;$
\begin{enumerate}
  \item (a) Each $\bm{W}$ is a stochastic real symmetric $n\times n$ matrix with $w_{ii}=0$. (b) $\bm{S}_1$ and $\bm{S}_2$ are non-singular for all $n$. (c) The sequences $\{\bm{W}\}$, $\{\bm{S}_1^{-1}\}$ and $\{\bm{S}_2^{-1}\}$
   are uniformly bounded in both row and column sums. (d) The largest eigenvalue of each $\bm{W}$ is bounded, $\max_i\lambda_i<\infty$.
  \item (a) The $n\times q$ instrument matrix $\bm{Z}_2$ and the $n\times (k_1 +1) $ matrix $[\bm{X}_1,\bm{x}_2]$ both have full column rank (for a large enough $n$), $\E[\bm{X}_1'\bm{\varepsilon}]=0$ and $\E[\bm{Z}_2'\bm{\varepsilon}]=0$ and (b) all the elements of $\bm{Z}_2$, $\bm{x}_2$ and $\bm{X}_1$ are uniformly bounded in absolute value.
  \item The innovations $\{\varepsilon_i:1\leq i \leq n, n\geq 1\}$ are identically distributed triangular arrays.
   Further the innovations $\{\varepsilon_i:1\leq i \leq n\}$ are for each n distributed (jointly) independently
  with $\E[\bm{\varepsilon}]=0$, $\E[\varepsilon_i^2]=\sigma^2_{\varepsilon}\in (0,\infty)$ and $\E[\varepsilon_iu_{2,i}]=\sigma_{\varepsilon ,u} \neq 0$.
  \end{enumerate}
\end{assump}

Assumption \ref{as:spatial2}.1 is standard in the spatial econometrics literature \citep{kelpru98,kp99, lee04}. Note, assumption \ref{as:spatial2}.1 (a) is required for the spectral decomposition and Assumption \ref{as:spatial2}.1 (d) ensures that the elements of the eigenvectors have the same dependence coefficient as the elements of the SWM. Assumption \ref{as:spatial2}.2 and \ref{as:spatial2}.3 are standard assumptions in the instrument variables literature.

\begin{assump}[Sparse Spectral Representation] \label{as:esf2}
  $\;$
  \begin{enumerate}
    \item $f(\bm{W,y,X}_1)= \bm{E}\bm{\gamma}_{y,0} =\bm{E}_{\Omega_y}\bm{\gamma}_{\Omega_y}+ \pi_x$ and $g(\bm{W},\bm{x}_2, \bm{X}_1,\bm{Z}_2)=\bm{E}\bm{\gamma}_{x,0}=\bm{E}_{\Omega_x}\bm{\gamma}_{\Omega_x}+\pi_y$
     where $\pi_y$ and $\pi_x$ are approximation errors, $\bm{E}_{\Omega_y}$ and $\bm{E}_{\Omega_x}$ are $n\times s_2$ and $n\times s_1$ matrices with columns that correspond to the active sets $\Omega_y:=\supp(\bm{\gamma}_{y,0})$ and $\Omega_x:=\supp(\bm{\gamma}_{x,0})$, and $\bm{\gamma}_{\Omega_y}$ and $\bm{\gamma}_{\Omega_x}$ the corresponding vectors of unknown constants.
     \item $|\Omega|=s<n-k_1-q$ where $\Omega=\Omega_y\cup \Omega_x$
     \item  $\pi_x=O_p(n^{-\frac{1}{2}-c})$ and $\pi_y=O_p(n^{-\frac{1}{2}-c})$ with $c>0$ constant.
  \end{enumerate}
\end{assump}

The second set of assumptions relates to the ESF approximation itself. Assumption \ref{as:esf2}.1 says there exists a set of linearly dependent eigenvectors and corresponding unknown constants that will approximate the functions $f(\bm{W,y,X}_1)$ and $g(\bm{W},\bm{x}_2, \bm{X}_1,\bm{Z}_2)$. Assumption \ref{as:esf2}.2 assumes this approximation is weakly sparse and Assumption \ref{as:esf2}.3 assumes the approximation errors go to zero at a sufficient speed. 

Under these assumptions, the high-dimensionality ESF system (\ref{esf}) and (\ref{xdgpE}) can be expressed as the following low dimensional reduced form system of equations:\footnote{While these assumptions cannot be verified in practice or even in simulations, they are common feature in the ESF literature, as well as related methodology such as factor or principal component analysis}
\begin{align}
    \bm{y} &=  \bm{G}_{\Omega} \bm{\Upsilon}_{\Omega} +\bm{\varepsilon}_\pi \label{yesfsupp} \\
    \bm{x}_2 & =  \bm{Z}_{\Omega}\bm{\zeta}_{\Omega} + \bm{u}_\pi \label{xesfsupp}
\end{align}
where $\bm{G}_{\Omega}=[\bm{X}_1,\bm{x}_2,\bm{E}_{\Omega}]$, $\bm{Z}_{\Omega}=[\bm{X}_1,\bm{Z}_2, \bm{E}_{\Omega}]$ ,
 $\bm{\Upsilon}_{\Omega}=[\bm{\beta}_{1,0}', \beta_{2,0},  \bm{\gamma}_{\Omega}']'$,
$\bm{\zeta}_{\Omega}=[\bm{\zeta}_{1,0}', \bm{\zeta}_{2,0}', \bm{\zeta}_{3,\Omega}']'$, $\bm{\varepsilon}_\pi=\bm{\varepsilon}+\pi_y$ and $\bm{u}_\pi=\bm{u}_2+\pi_x$ .

Even assuming that the subset of relevant vectors $\Omega$ is known, establishing that \eqref{yesfsupp}-\eqref{xesfsupp} can be estimated by 2SLS is non-trivial, for two reasons. First, we have the two additional approximation errors $\pi_y$ and $\pi_x$ in the first and second stage errors and second, the standard weak law of large numbers (LLN) and central limit theorem for triangular arrays used for spatial models requires assuming the row-wise independence. This is not realistic here as $\bm{G}_{\Omega}$ and $\bm{Z}_{\Omega}$ contain elements of $\bm{E}$, constructed from a linear transformation of $\bm{W}$, a matrix which itself encapsulates the spatial dependence of observations. Establishing the theoretical properties of the procedure therefore requires formalising this dependence and applying the appropriate limit theorems.

To do so, we use the notion of $\psi$-dependence first proposed by \citet{dl99} for time-series data and adapted by \citet{LT_KMS20} to allow for cross-sectional dependence. This allows us to use the limit theorems proposed by \citet{LT_KMS20}. Roughly speaking, $\psi$-dependence measures the strength of dependence between two sets of random variables by the covariance of non-linear functions of the random variables.

Let $\{w_{ij},1\leq i\leq n,n\geq 1\}$, $j=1,\ldots,n$, be a triangular array of random variables, where $w_{ij}=w_{ij,n}$ denotes the $i,j$th element of matrix $\bm{W}$ which is derived from a spatial structure as follows. For any $a\in \N$, we endow $\R^{a}$ with distance:
\[
\mathtt{d}_a(\bm{q},\bm{h}) = \sum_{l=1}^a|q_l-h_l|
\]
where $\bm{q}=(q_1,\ldots, q_a)$ and $\bm{h}=(h_1,\ldots,h_a)$ are points in $\R^{a}$. Let $\pazocal{L}_{a}$ denote the family of real valued, bounded Lipschitz functions, with $\lip(f)$ the Lipschitz constant of $f$ and $||f||_\infty=\sup_x|f(x)|$ its sup-norm.
\[
\pazocal{L}_{a}=\{f:\R^{a}\to \R:||f||_\infty <\infty ;\; \lip(f)<\infty\}
\]

Now consider two sets of cross-sectional units (of size $a$ and $b$ $\in\N$) with a distance between each other of at least $r>0$. Let $\pazocal{P}_{a,b;r}$ denote the collections of all pairs
\[
\pazocal{P}_{a,b;r} = \{(A,B):A,B\subset N, |A|=a,\; |B|=b,\; d_{A,B}\geq r\}
\]
where $d_{A,B}=\min_{i\in A}\min_{j\in B}d_{ij}$.\footnote{Note that $\pazocal{P}_{a,b;r}$, $d_{A,B}$ and $d_{ij}$ are also implicitly indexed by $n$, but we again omit the index to simplify the notation} $\forall$ sets $A$ of positive integers, define $w_A=\{w_{ij}:i\in A\}$.

We take $\{\C_n=\C\}$ be a sequence of given $\sigma$-fields, such that for each $n\geq 1$, the spatial weights matrix $\bm{W}_n=\bm{W}$ is $\C$-measurable. Definition \ref{def:psi} gives the exact definition of conditional $\psi$ dependence we use.

\begin{defin}[ ]\label{def:psi}
  \citep{LT_KMS20} The triangular array $\{w_{ij,n}=w_{ij},1\leq i\leq n,n\geq 1\}$, $j=1,\ldots,n$ is called conditionally $\psi$-dependent given $\{\C_n=\C\}$, if for each $n\in\N$ there exists a $\C$-measurable sequence $\mu_{r}=\{\mu_{r}=\mu_{r,n}:r\geq 0\}$,
  $\mu_{0}=1$, and a collection of non-random functions $\psi_{a,b}: \pazocal{L}_{a} \times \pazocal{L}_{b} \to [0,\infty)$ such that for all $(A, B) \in \pazocal{P}_{a,b;r}$ with $r > 0$ and all $f \in \pazocal{L}_{a}$ and  $g \in \pazocal{L}_{b}$,
 \begin{equation}\label{covmix}
   \big|\cov(f(w_A),g(w_B)|\C)\big| \enspace \leq \enspace \psi_{a,b}(f,g)\mu_{r} \quad \as
 \end{equation}
The sequence $\{\mu_{r}\}$ is the dependence coefficients of $\{w_{ij}\}$
\end{defin}

We will now explicitly specify the latent spatial formation process. We consider binary connectivity based on physical distance plus some stochastic elements. Specifically, the connection for each pair of spatial units $i$ and $j$ ($i\neq j$) is randomly realised if and only if:
\begin{equation*}
    w_{ij}=\mathbbm{1}\{ \phi_{ij} \geq \eta_{ij} \}
\end{equation*}
where the $\phi_{ij}$'s and $\eta_{ij}$'s are random variables such that $\phi_{ij,n}=\phi_{ij}=\phi_{ji}$, $\eta_{ij}=\eta_{ji}$ and $\{\eta_{ij} \; : \; i < j \}$ is $i.i.d.$ and independent of $\phi = (\phi_{ij})_{i<j}$. The random variable $\phi_{ij}$ which determines the formation probabilities is assumed to be a function of observable characteristics $\bm{l}_{ij,n}=\bm{l}_{ij}$(e.g., the physical distance between the spatial units) and unit specific unobservable characteristics $\bm{t}_{i,n}=\bm{t}_i$ (i.e. $\phi_{ij}=f(\bm{t}_i, \bm{t}_j, \bm{l}_{ij})$ where $f(\cdot )$ is some function). Thus, the $\sigma$-field $\C$ is generated by $\bm{t}_i$, $\bm{t}_j$ and $\bm{l}_{ij}$ for all $i$ and $j$.

Let us introduce following additional notations. Let  $\tilde{\bm{E}}$ be either equal to $\bm{M}_H \bm{E}$ or $\bm{M}_{\hat{\bm{X}}} \bm{E}$, and $\bm{C}_{\Omega k\Omega k}=n^{-1}\tilde{\bm{E}}_{\Omega k}'\tilde{\bm{E}}_{\Omega k}$, $\bm{C}_{\Omega k\grave{\Omega k}}=n^{-1}\tilde{\bm{E}}_{\Omega k}'\tilde{\bm{E}}_{\grave{\Omega k}}$, $\bm{C}_{\grave{\Omega k}\Omega k}=n^{-1}\tilde{\bm{E}}_{\grave{\Omega k}}'\tilde{\bm{E}}_{\Omega k}$ and $\bm{C}_{\grave{\Omega k}\grave{\Omega k}}=n^{-1}\tilde{\bm{E}}_{\grave{\Omega k}}'\tilde{\bm{E}}_{\grave{\Omega k}}$
where $\tilde{\bm{E}}_{\Omega k}$ is an $n\times s_k$ matrix with columns corresponding to the active set $\Omega k$. $\grave{\Omega k}$ is the complement set and the $n\times q_k$ matrix $\tilde{\bm{E}}_{\grave{\Omega k}}$ is defined accordingly with $q_{nk}=q_k=s_k-n$. Now the (re-scaled) Gram matrix $\bm{C}_n=\bm{C}=n^{-1}\tilde{\bm{E}}'\tilde{\bm{E}}$ can be expressed in block-wise form as:
\begin{equation*}
 \bm{C}  = \begin{bmatrix} \bm{C}_{\Omega k\Omega k} & \bm{C}_{\Omega k\grave{\Omega k}} \\   \bm{C}_{\grave{\Omega k}\Omega k} & \bm{C}_{\grave{\Omega k}\grave{\Omega k}} \end{bmatrix}.
\end{equation*}
Similarly we define $\bm{\gamma}=[\bm{\gamma}_{\Omega k},\bm{\gamma}_{\grave{\Omega k}}]'=[\gamma_1,\ldots,\gamma_{s_k},\gamma_{s_k+1},\ldots,\gamma_{n}]'$, with $k=x$ or $y$

\begin{assump}[Selection Consistency]\label{as:hidem}
    There exists $M_1,M_2,M_3 > 0$, $0\leq c_1<c_2 \leq 1$ and a vector of positive constants $\bm{\nu}$, the following holds:
  \begin{enumerate}
    \item \qquad $\frac{1}{n}\tilde{\bm{e}}_{i}'\tilde{\bm{e}}_{i} \leq M_1 \;\; \forall i,$

    \item \qquad $\bm{\alpha}'\bm{C}_{\Omega k\Omega k}\bm{\alpha} \geq M_2 \;\; \forall \; ||\bm{\alpha}||_2^2=1,$ with $k=x$ or $y$,
    \item \qquad $n^{\frac{1-c_2}{2}}\min_{i=1,\ldots,s_{k}}|\gamma_{i}|\geq M_3,$, with $k=x$ or $y$
    \item \qquad $s_k = O(n^{c_1}),$ with $k=x$ or $y$,
    \item \qquad $ |\bm{C}_{\grave{\Omega k}\Omega k}(\bm{C}_{\Omega k\Omega k})^{-1}\sign(\bm{\gamma}_{\Omega k})|\leq \bm{1}-\bm{\nu},$ with $k=x$ or $y$.
  \end{enumerate}
\end{assump}

Assumption \ref{as:hidem} is similar to  Assumption 4 in \citet{c23}. These are conditions on the eigenvectors and eigenvalues to assure consistent selection.

\subsection{Consistent Eigenvector Selection}

We now derive conditions under which Algorithm \ref{Mi-2SLalg} selects the relevant eigenvectors in steps 3 and 4. \citet{c23} discusses the conditions for consistent selection in a SAR model, which involve some restrictions on eigenvalues and the level of sparsity $s_1$ and $s_2.$

\begin{defin}\label{def:con} Mi-Lasso estimates of $\bm{\gamma}$ are selection consistent if:
\begin{equation*}
 \lim_{n\to\infty}P(\tilde{\bm{\gamma}}=_s \bm{\gamma}_0) =1.
\end{equation*}
\end{defin}

\begin{lem}\label{lem:selcon}
  Assuming Assumption \ref{as:spatial2}, \ref{as:esf2} and \ref{as:hidem} hold, and $c_2-c_1=0.5$. Given $s_k+q_k=n$ implies Mi-Lasso in Algorithm \ref{Mi-2SLalg} at the steps 3 and 4 are sign consistent for all $\frac{1}{z_k^2}$ that satisfy $\frac{1}{z_k^2\sqrt{n}}= o_p(n^{\frac{c_2-c_1}{2}})= o_p(n^{\frac{1}{4}})$ and $\frac{1}{n^3z_k^{8}}\to \infty$, with $k=x$ or $y$
  we have:
  $$\p\big(\hat{\bm{\gamma_k}}=_s\bm{\gamma}_{0k}\big)\geq 1 -O(n^3z_k^{8}) \to 1 \;\;\; as \; n\to \infty,$$ with $k=x$ or $y.$
\end{lem}

\textbf{Proof:} The proof of the Lemma \ref{lem:selcon} follows immediately form the application of Theorem 2 from \citet{c23}.

\subsection{Estimation consistency}

We will now derive a consistency proof for estimating $\bm{\Upsilon}_{\Omega}$ by 2SLS, assuming $\Omega$ is known. In scalar notation \eqref{yesfsupp} can be rewritten as:
\begin{align*}
  y_i & = \sum^{k_1}_{j=1}x_{ij,1}\beta_{j,1,0} + x_{i,2}\beta_{2,0} + \sum^{s}_{j=1}e_{ij}\gamma_{j,\Omega} + \varepsilon_i = \sum^{(k_1+1+s)}_{j=1}g_{ij,\Omega}\Upsilon_{j,\Omega} + \varepsilon_{i\pi}\\
  x_{i,2} & = \sum^{k_1}_{j=1}x_{ij,1} \zeta_{j,1,0} + \sum^q_{j=1}z_{ij,2}\zeta_{j,2,0} + \sum^{s}_{j=1}e_{ij}\zeta_{j,3,\Omega} + u_{i,2}= \sum^{(k_1+q+s)}_{j=1}z_{ij,\Omega}\psi_{j,\Omega} + u_{i,2\pi} \\
\end{align*}
for $i=1,\ldots, n$.

We now state the additional assumptions for consistent estimation of $\beta_{2,0}$ by 2SLS.

\begin{assump}[LLN restrictions on conditional $\psi$-dependence]\label{as:lln2}
  $\;$
  \begin{enumerate}
    \item The triangular array $\{w_{ij}\}$, is conditionally $\psi$-dependent given $\{\C\}$ with the dependence coefficients $\{\mu_{r}\}$ satisfying the following condition. For some constant $C > 0$
 \begin{equation}\label{psi_con}
   \psi_{a,b}(f,g) \enspace \leq \enspace C ab(||f||_\infty+\lip(f))(||g||_\infty+\lip(g))
 \end{equation}
 \item For some $l>2$:
 \begin{align*}
  \sup_{n\geq 1} \max_{i\in N}\left(\E\left[|y_{i}|^{l}|\C\right]\right)^{1/l} & \enspace < \enspace \infty \quad \as, \\
  \sup_{n\geq 1} \max_{i\in N}\left(\E\left[\sum^{(k_1+1+s)}_{j=1}|g_{ij,\Omega}|^{l}|\C\right]\right)^{1/l} & \enspace < \enspace \infty \quad \as \\
 \sup_{n\geq 1} \max_{i\in N}\left(\E\left[\sum^{(k_1+q+s)}_{j=1}|z_{ij,\Omega}|^{l}|\C\right]\right)^{1/l} & \enspace < \enspace \infty \quad \as
\end{align*}
 \item
 \begin{equation}\label{dencon}
   n^{-1}\sum_{r=1}^\infty \delta^d_r\mu_{r} \enspace \to_{a.s.} \enspace 0, \quad n \to \infty
 \end{equation}
 where $\delta^d_r=n^{-1}\sum_{i\in N}|N^d_{i,r}|$ and $N^d_{i,r}=\{j\in N:d_{i,j}=r\}$ denotes the set of cross-sectional units exactly distance $r$ from unit $i$.
  \item $\E[\bm{Z}_{\Omega}'\bm{\varepsilon}| \C] \enspace = \enspace 0$.
  \end{enumerate}
\end{assump}

Assumption \ref{as:lln2}.1 is from \citet{LT_KMS20} and the function $\psi_{a,b}$ satisfies Assumption \ref{as:lln2}.1 if:
\[
\sup_{n\geq 1}\max_{i\in N}\E[|w_{ij}|^q|\C_n] \enspace < \enspace \infty \quad \as
\]
for some $q>4$ and $\forall j$. Assumption \ref{as:lln2}.2 states that all variables have conditional finite second moments, so all are $\C$ measurable. Assumption \ref{as:lln2}.3 is also from \citet{LT_KMS20} and puts a restriction on the denseness of the spatial structure and the rate of decay of dependence with regards to the distance between the spatial units. In the mixing literature, it is common to assume the mixing coefficients can be summed $n^{-1}\sum_{r=1}^\infty \mu_{r}=O_p(1)$ as $n\to\infty$. A sufficient condition for Assumption \ref{as:lln2}.3, in this case, is if the average number of neighbours at distance $r$ grows slower than the sample size $n$, i.e. $\sup_{r\geq 1} \delta^d_r = o_p(n)$.
Intuitively, this assumption requires that the number of spatial connections at distance $r$ not grow too fast as $r$ increases. However, as the precise condition \eqref{dencon} includes the dependence coefficient $\mu_{r}$, this assumption can be relaxed if $\mu_{r}$ itself decreases at an appropriate rate relative to $r$. This assumption seems reasonable as the literature on estimating SWMs often assumes a sparse spatial structure \citep{arhbha15,LS16_SWM,LS20_SWM}. An example of where Assumption \ref{as:lln2}.3 could fail is if one unit is connected to all other units, such as in the star network. This is because the distance between any two units is never larger than 2.\footnote{$\delta^d_1 = 2(n-1)/n$, $\delta^d_2 = (n-2)(n-1)/n$ and $\delta^d_r = 0$ for $r\geq 3$}
Assumption \ref{as:lln2}.4 requires the instruments (including $\bm{E}_\Omega$) be uncorrelated with the structural error, conditional on $\C$.

Lemma \ref{eigtran} below establishes that assumption \ref{as:lln2}.1, which requires that $w_{ij}$ are $\psi$-dependent triangular arrays, carries over to the eigenvector elements $e_{ik}$. Given the eigendecomposition $\bm{W} = \bm{E} \bm{\Lambda} \bm{E}^T$, these are generated by a linear combination of $w_{ij}$, $\lambda_k$ and $e_{jk}$ as follows:
\begin{equation}
  \label{eigenvec_elements}
  e_{ik} = \sum^n_{j=1} w_{ij}e_{jk}/\lambda_k
\end{equation}
for all $\lambda_k \neq 0$
\medskip

\begin{lem}\label{eigtran}
 Suppose the triangular array $\{w_{ij}\}$, with $w_{ij} \in\R$ satisfies Assumption \ref{as:lln2}.1 with dependence coefficient $\{\mu_r\}$. For each $n \geq 1$ let $\{\lambda_{k,n}=\lambda_k\}_{k\in N}$, $\lambda_k\in \R $, $\lambda_k \neq 0$
  and $\{\bm{e}_{k,n}=\bm{e}_{k}\}_{k\in N}$, $\bm{e}_{k}\in\R^n$ be a sequence of $\C$ measurable random scalars and random vectors with $\max_{k\in N} |\lambda_k| \leq \infty$ a.s. and $ ||\bm{e}_{k}||_2^2 = 1 \; \forall k$.
 Then the array $\{e_{ik}\}$ defined by \eqref{eigenvec_elements} for $i=1,\ldots, n$ and $k=1,\ldots, n$ is conditionally $\psi$-dependent given $\{\C\}$ with the dependence coefficients $\{\mu_r\}$,
 \begin{equation*}
   \left|\cov\left(f\left(\sum_{j\in N} w_{a}e_{jk}/\lambda_k\right),g \left(\sum_{j\in N}w_{b}e_{jk}/\lambda_k\right)|\C\right)\right| \enspace \leq \enspace \psi_{a,b}(f_c,g_c)\mu_{r} \quad \as
 \end{equation*}
\end{lem}

\textbf{Proof:} This is provided in appendix \ref{app:proofs2}.

Lemma \ref{eigtran} shows that as long as the largest eigenvalue is bounded and the eigenvectors are mutually orthogonal (both of these requirements are satisfied by Assumption \ref{as:spatial2}.1) the eigenvector elements will have the same dependence coefficients $\{\mu_r\}$ as the elements of the SWM.

\begin{theorem}\label{theo:IVsfplim}
Assuming Assumption \ref{as:spatial2}-\ref{as:lln2} holds we have:
\begin{equation*}
\hat{\bm{\Upsilon}}_{\Omega} \to_{p} \bm{\Upsilon}_{\Omega}
\end{equation*}
where $\hat{\bm{\Upsilon}}_{\Omega}$ is the estimate of $\bm{\Upsilon}_{\Omega}$ from \eqref{yesfsupp} obtained using Algorithm \ref{Mi-2SLalg}.
\end{theorem}

\textbf{Proof:} This is provided in appendix \ref{app:proofs2}.

Theorem \ref{theo:IVsfplim} shows that under an appropriate mixing condition, some additional regularity conditions and if $\Omega$ is known, we could estimate $\bm{\Upsilon}_{\Omega}$ consistently by 2SLS. The proof of Theorem \ref{theo:IVsfplim} uses the weak LLN for triangular arrays, which gives convergence in probability, and the strong LLN for cross-sectionally dependent random variables of \citet{LT_KMS20} which gives almost sure convergence, thus, overall gives convergence in probability. An almost sure convergence result could be obtained similarly by using the strong LLN for triangular arrays instead of the weak LLN for triangular arrays.

\subsection{Asymptotic Distribution }

In order to derive the asymptotic distribution of the 2SLS estimator for a known support $\Omega$ of the relevant eigenvector set, we need some additional assumptions:

\begin{assump}[CLT restrictions on conditional $\psi$-dependence]\label{as:clt2}
  $\;$
  \begin{enumerate}
    \item for some $l>4$:
    \begin{align*}
      \sup_{n\geq 1} \max_{i\in N}\left(\E[|y_{i}|^{l}|\C]\right)^{1/l} & \enspace < \enspace \infty,  \\
      \sup_{n\geq 1} \max_{i\in N}\left(\E\left[\sum^{(k_1+1+s)}_{j=1}|g_{ij,\Omega}|^{l}|\C\right]\right)^{1/l} & \enspace < \enspace \infty  \quad \as \textrm{ and} \\
      \sup_{n\geq 1} \max_{i\in N}\left(\E\left[\sum^{(k_1+q+s)}_{j=1}|z_{ij,\Omega}|^{l}|\C\right]\right)^{1/l} & \enspace < \enspace \infty
    \end{align*}
     \item There exists a positive sequence $m_n = m \to \infty$ such that for $k = 1, 2$
 \begin{align}
   n \bm{\Sigma}^{-(2+k)} \sum_{r=0}^\infty c_{r,m;k}\mu_r^{1-\frac{2+k}{l}} \to_{a.s.} 0, \\
   n^2\mu^{1-{1/l}}_{m}\bm{\Sigma}^{-1} \to_{a.s.}0, \label{decay}
 \end{align}
 where $\bm{\Sigma}=\E[\bm{Z}_\Omega'\bm{Z}_\Omega |\C]\sigma^2_\varepsilon$,  $c_{r,m;k}=\inf_{\alpha>1}\big[\Delta_{r,m;k\alpha}\big]^{1/\alpha}\big[\delta^d_{r,\alpha/(1-\alpha)}\big]^{1-1/\alpha}$,

 $\delta^d_{r,k}=n^{-1}\sum_{i\in N}|N^d_{i,r}|^k$,
 $\Delta_{r,m;k}=n^{-1}\sum_{i\in N}\max_{j\in N^d_{i,r}} |N_{i,m}/ N_{j,r-1}|^k$, $N_{i,r}=\{j\in N:d_{i,j}\leq r\}$,  $N^d_{i,r}=\{j\in N:d_{i,j}=r\}$
 and $l>4$ is as same as in Assumption \ref{as:clt2}.1. As $n\to\infty$.
  \end{enumerate}
\end{assump}

Assumption \ref{as:clt2}.1 states that all variables have at least conditional fourth finite moment, so are all $\C$ measurable, which is in line with many spatial and 2SLS models. Assumption \ref{as:clt2}.2 is from \citet{LT_KMS20} and limits the extent of the spatial dependence of the random variables through restrictions on the spatial structure. When the spatial structure is given $c_{r,m;k}$ can be computed, it is composed of two parts $\Delta_{r,m;k\alpha}$ and $\delta^d_{r,\alpha/(1-\alpha)}$, which capture the denseness of the spatial structure through the average size of neighbourhoods and the average shell size of the neighbourhood. Note that after $r$ goes beyond a certain level $\Delta_{r,m;k}$ tends to decrease fast, as the set $N_{j,r-1}$ becomes large quickly. For \eqref{decay} to be satisfied $\mu_r$ (the spatial dependence) needs to decay fast enough as $r$ becomes large, this is because it will become increasingly difficult to find a slowly increasing sequence $m$ to satisfy the condition.

\begin{theorem}\label{theo:IVsfdist}
Assuming Assumptions \ref{as:spatial2}-\ref{as:clt2} holds we have
\begin{equation*}
\sqrt{n}(\hat{\bm{\Upsilon}}_{\Omega}-\bm{\Upsilon}_{\Omega}) \to_d N(0,n \left(plim_{n \to \infty}\big([\bm{G}_{\Omega}'\bm{Z}_{\Omega}|\C][\bm{Z}_{\Omega}'\bm{Z}_{\Omega}|\C]^{-1}
      [\bm{Z}_{\Omega}'\bm{G}_{\Omega}|\C]\big)^{-1} \right)\sigma_\varepsilon^2 )
\end{equation*}
where $\hat{\bm{\Upsilon}}_{\Omega}$ is the estimate of $\bm{\Upsilon}_{\Omega}$ from \eqref{yesfsupp} obtained using the Algorithm 1 Mi-2SL.
\end{theorem}

\textbf{Proof:} This is provided in appendix \ref{app:proofs2}.

Theorem \ref{theo:IVsfdist} shows that if $\Omega$ is known, then under an appropriate mixing condition, restriction on the denseness of the spatial structure, and some additional regularity conditions, the 2SLS estimate of $\bm{\Upsilon}_{\Omega}$ and thus, $\beta_{2,0}$ will be asymptotically normal, with a  convergence rate of $n^{-1/2}$.

\section{Simulation}\label{sims2}

In this section, we provide simulation evidence to assess the finite sample performance of the Mi-2SL estimator and compare its performance to some commonly used estimator for spatial models. We generate the following system of equations \eqref{simdgpa} - \eqref{simdgpb} where the structural equation includes a SAR(1) with spatial lags of the exogenous/endogenous variables, and the endogenous variable follows a SAR(2) with spatial lags of the exogenous variable/instrument:
\begin{align}
  \bm{y}=&\bm{Wy}\rho + \beta_1\bm{x}_1+\beta_2\bm{x}_2+ \bm{Wx}_1\omega+\bm{Wx}_2\omega +\bm{u} \label{simdgpa} \\
  \bm{x}_2=&  \zeta_1\bm{x}_1 + \zeta_2\bm{z}_2+ \sum_{i=1}^2\bm{W}^i\bm{x}_2\zeta_{3,i} + \bm{Wx}_1\omega+ \bm{Wz}_2\omega +\bm{v}  \label{simdgpb}
\end{align}
with $\bm{z}_2\sim N(0,\bm{I})$, $\bm{x}_1\sim  N(0,\bm{I})$ and $u_i$, $v_i$ (the $i$th elements of $\bm{u}$ and $\bm{v}$) are given by:
\[
  (u_i,v_i)\sim N\left(0,
  \begin{pmatrix}
    1 & \sigma_{v,u} \\
    \sigma_{v,u} & 1
  \end{pmatrix}\right)
\]

We set the non-spatial parameters to $\zeta_1=\zeta_2=\beta_1=\beta_2=1$ and $\sigma^2_{v,u}=0.9$, and the spatial parameters are combinations of the following values: $\rho\in\{0,0.4,0.8\}$, $\zeta_{3,1}\in\{0.4,0.8\}$, $\zeta_{3,1}\in\{0,0.4\}$ and $\omega\in\{0,0.4,0.8\}$.

The SWM $\bm{W}$ is generated using a \citet{watts1998collective} small world network model. Small world networks are a popular way of modelling cross-sectional dependency in social networks, and have been used in many economic applications, particularly the economics of innovation diffusion and industrial clusters \citep[see for example][]{jackson2005economics, cassi2008opportunity, maggioni2011networks, ter2011co, gulati2012rise, bagley2019small}. The number of neighbours is set to 10 and the rewiring probabilities to $p\in\{0.4,0.8\}$. This allows to see the difference been a higher level of clustering ($p=0.4$) and a lower level of clustering ($p=0.8$). Each SWM is normalised by the largest row sum and the eigenvectors are from the normalised SWM. Sample sizes considered are $n\in\{100,250,500\}$, and we run 1000 Monte Carlo replications for each experiment.

The estimators and specifications compared are:
\begin{enumerate}
  \item Naïve OLS (denoted simpOLS). This estimates $\bm{y}=\alpha\bm{\iota}+\beta_1\bm{x}_1+\beta_2\bm{x}_2+\bm{e}$ by OLS, ignoring the spatial process and the endogeneity of $\bm{x}_2$.
  \item Naïve IV (denoted simpIV). This estimates $\bm{y}=\beta_1\bm{x}_1+\beta_2\bm{x}_2+\bm{e}$ by IV with $\bm{z}_2$ as instrument for $\bm{x}_2$, but ignores the spatial process.
  \item 2SLS with a SAR(1) in the equation \eqref{simdgpa} (denoted 2SLS-SAR). This estimates $\bm{y}=\rho\bm{Wy}+\bm{x}_1\beta_1+\bm{x}_2\beta_2 +\bm{e}$ by 2SLS, with $\sum_{i=1}^{2}\bm{W}^{i}\bm{x}_1$ and $\bm{z}_2$ as instruments for $\bm{Wy}$ and $\bm{x}_2$, but ignoring spatial lags of the covariates in \eqref{simdgpb}.
  \item The Mi-2SL Algorithm \ref{Mi-2SLalg} with the first-stage fitted values from Lasso (step 3) (denoted Mi-2SLl).
  \item The Mi-2SL Algorithm \ref{Mi-2SLalg} with the first-stage fitted values from post-Lasso (step 3) (denoted Mi-2SLpl).
\end{enumerate}

\setlength{\tabcolsep}{5pt} 

\begin{table}[!h]
  \centering
  \caption{Results for $n=100$ and $\omega=0.4$}
  \label{tab:n100,o4}
  \begin{threeparttable}
    {\footnotesize
    \begin{tabular}{ccclrcccrccc}
      \hline
      \multicolumn{3}{c}{Experiment} & \T \B & \multicolumn{4}{c}{Rewiring prob. $p=0.4$} & \multicolumn{4}{c}{Rewiring prob. $p=0.8$} \\
      \T \B $\rho$ & $\zeta_{3,1}$ & $\zeta_{3,2}$ & Estimator & bias & MSE & AASE & Vecs & bias & MSE & AASE  &  Vecs \\
       \hline
       \multirow{5}{*}{0.4} & \multirow{5}{*}{0.4} & \multirow{5}{*}{  0} & \T  SimpOLS &  0.490 &  0.243 &  0.058 & - &  0.504 &  0.258 &  0.059 & - \\
                            &                      &                      &      SimpIV &  0.007 &  0.013 &  0.111 & - &  0.019 &  0.013 &  0.112 & - \\
                            &                      &                      &    2SLS-SAR & -0.012 &  0.012 &  0.107 & - & -0.007 &  0.011 &  0.107 & - \\
                            &                      &                      &     Mi-2SLl & -0.012 &  0.018 &  0.090 & [2,19] 20 & -0.004 &  0.019 &  0.089 & [2,21] 22 \\
                            &                      &                      & \B Mi-2SLpl & -0.010 &  0.018 &  0.093 & [2,15] 16 & -0.004 &  0.019 &  0.092 & [2,16] 18 \\

       \multirow{5}{*}{0.4} & \multirow{5}{*}{0.4} & \multirow{5}{*}{0.4} & \T  SimpOLS &  0.486 &  0.240 &  0.058 & - &  0.500 &  0.254 &  0.058 & - \\
                            &                      &                      &      SimpIV &  0.008 &  0.013 &  0.110 & - &  0.020 &  0.013 &  0.111 & - \\
                            &                      &                      &    2SLS-SAR & -0.012 &  0.012 &  0.106 & - & -0.008 &  0.011 &  0.106 & - \\
                            &                      &                      &     Mi-2SLl & -0.012 &  0.018 &  0.088 & [3,20] 21 & -0.005 &  0.020 &  0.087 & [3,22] 23 \\
                            &                      &                      & \B Mi-2SLpl & -0.010 &  0.017 &  0.091 & [3,14] 17 & -0.003 &  0.019 &  0.090 & [3,16] 19 \\

       \multirow{5}{*}{0.4} & \multirow{5}{*}{0.8} & \multirow{5}{*}{  0} & \T  SimpOLS &  0.500 &  0.254 &  0.056 & - &  0.517 &  0.271 &  0.055 & - \\
                            &                      &                      &      SimpIV &  0.017 &  0.013 &  0.110 & - &  0.030 &  0.013 &  0.111 & - \\
                            &                      &                      &    2SLS-SAR & -0.012 &  0.012 &  0.107 & - & -0.007 &  0.011 &  0.108 & - \\
                            &                      &                      &     Mi-2SLl & -0.006 &  0.020 &  0.088 & [10,16] 23 &  0.003 &  0.021 &  0.087 & [12,17] 26 \\
                            &                      &                      & \B Mi-2SLpl & -0.011 &  0.020 &  0.093 & [10,9] 18 &  0.000 &  0.020 &  0.092 & [12,9] 21 \\

       \multirow{5}{*}{0.4} & \multirow{5}{*}{0.8} & \multirow{5}{*}{0.4} & \T  SimpOLS &  0.497 &  0.251 &  0.055 & - &  0.516 &  0.270 &  0.054 & - \\
                            &                      &                      &      SimpIV &  0.018 &  0.013 &  0.109 & - &  0.032 &  0.013 &  0.110 & - \\
                            &                      &                      &    2SLS-SAR & -0.012 &  0.012 &  0.106 & - & -0.007 &  0.011 &  0.107 & - \\
                            &                      &                      &     Mi-2SLl & -0.008 &  0.019 &  0.087 & [13,14] 24 &  0.002 &  0.020 &  0.087 & [16,14] 27 \\
                            &                      &                      & \B Mi-2SLpl & -0.013 &  0.020 &  0.092 & [13,7] 20 & -0.001 &  0.020 &  0.092 & [16,7] 23 \\

       \multirow{5}{*}{0.8} & \multirow{5}{*}{0.4} & \multirow{5}{*}{  0} & \T  SimpOLS &  0.539 &  0.296 &  0.066 & - &  0.567 &  0.327 &  0.069 & - \\
                            &                      &                      &      SimpIV &  0.034 &  0.017 &  0.122 & - &  0.052 &  0.019 &  0.126 & - \\
                            &                      &                      &    2SLS-SAR & -0.015 &  0.012 &  0.107 & - & -0.010 &  0.011 &  0.108 & - \\
                            &                      &                      &     Mi-2SLl & -0.020 &  0.023 &  0.078 & [2,43] 44 & -0.009 &  0.024 &  0.077 & [2,48] 48 \\
                            &                      &                      & \B Mi-2SLpl & -0.011 &  0.023 &  0.079 & [2,39] 41 & -0.002 &  0.024 &  0.079 & [2,43] 44 \\

       \multirow{5}{*}{0.8} & \multirow{5}{*}{0.4} & \multirow{5}{*}{0.4} & \T  SimpOLS &  0.538 &  0.295 &  0.066 & - &  0.568 &  0.328 &  0.068 & - \\
                            &                      &                      &      SimpIV &  0.035 &  0.017 &  0.121 & - &  0.054 &  0.019 &  0.126 & - \\
                            &                      &                      &    2SLS-SAR & -0.015 &  0.012 &  0.106 & - & -0.011 &  0.011 &  0.107 & - \\
                            &                      &                      &     Mi-2SLl & -0.015 &  0.023 &  0.076 & [3,45] 45 & -0.006 &  0.024 &  0.076 & [3,50] 50 \\
                            &                      &                      & \B Mi-2SLpl & -0.008 &  0.023 &  0.077 & [3,39] 41 &  0.001 &  0.024 &  0.078 & [3,43] 46 \\

       \multirow{5}{*}{0.8} & \multirow{5}{*}{0.8} & \multirow{5}{*}{  0} & \T  SimpOLS &  0.569 &  0.329 &  0.063 & - &  0.605 &  0.372 &  0.065 & - \\
                            &                      &                      &      SimpIV &  0.049 &  0.018 &  0.122 & - &  0.071 &  0.022 &  0.127 & - \\
                            &                      &                      &    2SLS-SAR & -0.014 &  0.012 &  0.108 & - & -0.009 &  0.011 &  0.108 & - \\
                            &                      &                      &     Mi-2SLl &  0.017 &  0.024 &  0.073 & [10,43] 47 &  0.032 &  0.028 &  0.075 & [12,46] 51 \\
                            &                      &                      & \B Mi-2SLpl &  0.013 &  0.024 &  0.081 & [10,29] 38 &  0.032 &  0.028 &  0.081 & [12,32] 42 \\

       \multirow{5}{*}{0.8} & \multirow{5}{*}{0.8} & \multirow{5}{*}{0.4} & \T  SimpOLS &  0.572 &  0.332 &  0.063 & - &  0.619 &  0.390 &  0.065 & - \\
                            &                      &                      &      SimpIV &  0.051 &  0.018 &  0.121 & - &  0.075 &  0.023 &  0.128 & - \\
                            &                      &                      &    2SLS-SAR & -0.014 &  0.012 &  0.107 & - & -0.009 &  0.011 &  0.108 & - \\
                            &                      &                      &     Mi-2SLl &  0.022 &  0.024 &  0.073 & [13,42] 48 &  0.037 &  0.028 &  0.075 & [16,43] 51 \\
                            &                      &                      & \B Mi-2SLpl &  0.014 &  0.023 &  0.081 & [13,27] 38 &  0.037 &  0.028 &  0.083 & [16,27] 41 \\
       \hline
     \end{tabular}
     }
  {\footnotesize
  \begin{tablenotes}
      \item Note: bias is the bias of $\beta_2$, MSE is the mean squared error of $\beta_2$, AASE is the average asymptotic standard error of $\beta_2$ and [a,b] c is the average number of eigenvectors selected/used in steps 3, 5 and 6 of Algorithm \ref{Mi-2SLalg}.
  \end{tablenotes}
  }
  \end{threeparttable}
\end{table}

\setlength{\tabcolsep}{6pt} 

Note, 2SLS-SAR is based on the procedure proposed by \citet{kelpru98} (also commonly refereed to as Generalised Spatial two Stage Least Squares). This is included in the comparison set as it is a common spatial model used by applied researchers, and provides a more challenging benchmark for Mi-2SL, as it is a specification where a genuine attempt is made at controlling for both the endogeneity of $\bm{x}_2$ and the presence of a spatial process.

Tables \ref{tab:n100,o4}, \ref{tab:n250,o4}, and \ref{tab:n500,o4} presents the bias, mean squared error (MSE) and average asymptotic standard error (AASE) of $\beta_2$ for $\omega=0.4$ and sample sizes of 100, 250 and 500 respectively.\footnote{Tables \ref{tab:n250,o4}, and \ref{tab:n500,o4} are provided in appendix \ref{sims2full}. Additional extended results for $\omega=\{0,0.8\}$ and the $\rho=0$ case can be found in the supplementary material} These tables exhibit the standard bias-variance trade off between naïve OLS and naïve IV. OLS has the smallest AASE but the largest bias, whereas IV eliminates a substantial part of the bias but has the largest AASE. As both ignore the presence of a spatial process, the 2SLS-SAR is able to decrease both the bias and AASE compared to IV.

The Mi-2SL estimators have the second smallest AASE (smaller than simpIV and 2SLS-SAR) overall, and when $\rho$ is large Mi-2SL has the same or slightly larger AASE as OLS while having a small bias similar to the SAR(1). Generally when the rewiring probability is small ($p=0.4$) Mi-2SL has a larger absolute bias compared to when the rewiring probability is large, regardless of the sample size. In terms of eigenvector selection behaviour, the number of selected eigenvectors increases with $\rho$ and as the sample size increases. When the sample size is small ($n=100$) more eigenvectors are selected when $p=0.8$, whereas for larger sample sizes  more eigenvectors are selected when $p=0.4$. More eigenvector are selected when the first stage fitted values come from Lasso estimate (Mi-2SLl), this is because more eigenvectors are selected in the second stage.

In summary Mi-2SL performs well compared to OLS, naïve IV and the SAR(1) estimated by 2SLS. It has a smaller AASE than Classical IV and the SAR(1). In particular, when the level of spatial correlation of the dependent variable in the structural equation is high, Mi-2SL AASE is similar to that of OLS. In terms of bias Mi-2SL generally performs better than both OLS and IV, and similarly to the SAR(1).

\section{Application on impact of migration on labour markets}\label{app_ck16}

This section revisits the empirical application of \citet{ck16}. Using an IV strategy to control for the endogeneity of labour market decisions, their main finding is `that low-skilled Mexican-born immigrants’ location choices respond strongly to changes in local labour demand, which helps equalize spatial differences in employment outcomes for low-skilled native workers. \citet{ck16} starts from the observation that over the Great Recession low-educated Mexican-born male immigrants were more mobile than their native counterparts. Given this observation, their aim was to test if location choice of migrants was being driven by local labour market conditions, leveraging the geographic variation in employment changes during the Great Recession as a natural experiment. Their argument rests on fact that changes in labour market conditions during the Great Recession can be approximately measured by changes in employment, as traditionally sticky-downwards wages were essentially fixed during that period. Thus, they look at the effect of changes in employment on population changes for 20 different demographic groups, split by gender (males and females), education (`high school or less' and `some college or more'), and location of birth (native-born, foreign-born, Mexican-born, and other foreign-born). The unit of observation is a metropolitan statistical area (MSA), 95 of which are included in their IV analysis. Their empirical specification is:
\begin{equation}
  \label{ck16_spec}
  \begin{aligned}
      \Delta pop_i &= \beta_0 + \beta_1 \Delta emp_i + \beta_2 mex_i + \beta_3 policy_i + \beta_4 287g_i + u_i  \\
      \Delta emp_i & = \psi_0 + \psi_1 \Delta bartik_i + \psi_2 mex_i + \psi_3 policy_i + \psi_4 287g_i + v_i
  \end{aligned}
\end{equation}
where $i$ indexes the MSA, $\Delta pop_i$ is the proportional change in working-age population from 2006–2010, $\Delta emp_i$ is the proportional change in employment from 2006–2010, $mex_i$ is the share of Mexicans-born population in 2000, $policy_i$ and $287g_i$ are both immigration policy controls, and $\Delta bartik_i$ is the `Bartik instrument' \citet{b91}, which predicts changes in local labour demand by assuming that in each industry national employment changes are proportionately allocated across cities, based on each cities initial industry composition of employment. For the reader's convenience, Table \ref{ck16rep} replicates their main IV results, Table 4 in \citet{ck16}. We have also added the first stage (full) F-statistic, so this can be compared to the partial F-statistic and give further insight into the actual impact of the Bartik instrument in their estimates. Table \ref{ck16rep} shows the full F-statistic is always smaller than the partial F-statistic, implying that in their specification, the Bartik, which is supposed to give the identification, is not helping the first stage.

\begin{table}[t]
  \centering
  \caption{Replication of main IV results \citep[Table 4 in ][]{ck16}}
  \label{ck16rep}
  \begin{threeparttable}
    {\footnotesize
    \begin{tabular}{lccccc}
      \hline
      \T  &&&&& Other \\
      \B  & All & Native-born & Foreign-born & Mexican-born & foreign-born \\
      \cline{2-6}
      \multicolumn{6}{l}{\qquad \emph{Panel A: Men, high school or less} \T\B }   \\
   \T  Change in log of group-specific &  0.223 &  0.007 &  0.402 &  0.992 & -0.675 \\
       employment                      & (0.166) & (0.09) & (0.409) & (0.468) & (0.278) \\
       First stage F-statistic         & 11.42 & 10.94 & 12.76 & 8.28 & 17.11 \\
   \B  Partial F-statistic (Bartik)    & 35.74 & 36.14 & 25.31 & 11.94 & 45.61 \\
       \multicolumn{6}{l}{\qquad \emph{Panel B: Men, some college or more} \T\B }   \\
   \T  Change in log of group-specific &  0.27 &  0.411 & -0.237 & -0.475 & -0.161 \\
       employment                      & (0.157) & (0.192) & (0.264) & (0.387) & (0.329) \\
       First stage F-statistic         & 7.19 & 6.74 & 13.07 & 14.59 & 13.02 \\
   \B  Partial F-statistic (Bartik)   & 23.89 & 21.9 & 37.76 & 31.79 & 36.89 \\
       \multicolumn{6}{l}{\qquad \emph{Panel C: Women, high school or less} \T\B }   \\
   \T  Change in log of group-specific &  0.145 & -0.405 &  0.272 & 1.811 & -0.979 \\
       employment                      & (0.168) & (0.287) & (0.504) & (0.665) & (0.556) \\
       First stage F-statistic         & 10.43 & 8.76 & 15.21 & 6.04 & 22.42 \\
   \B  Partial F-statistic (Bartik)    & 28.59 & 26.09 & 26.76 & 13.74 & 39.17 \\
       \multicolumn{6}{l}{\qquad \emph{Panel D: Women, some college or more} \T\B }   \\
   \T  Change in log of group-specific & -0.066 & -0.054 & -0.754 &  0.438 & -1.092 \\
       employment                      & (0.378) & (0.42) & (0.716) & (0.919) & (0.738) \\
       First stage F-statistic         & 1.53 & 1.56 & 3.39 & 7.75 & 3.49 \\
   \B  Partial F-statistic (Bartik)    & 5.85 & 5.58 & 12.97 & 27.33 & 13.12 \\
      \hline
     \end{tabular}
     }
  {\footnotesize
  \begin{tablenotes}
      \item \emph{Note}: Robust standard errors in parentheses. See \citet{ck16} Table 4 for further details.
  \end{tablenotes}
  }
  \end{threeparttable}
\end{table}

\begin{figure}[t]
  \caption{Metropolitan areas in \citet{ck16}}
  \label{metck16}
\includegraphics[width=14cm]{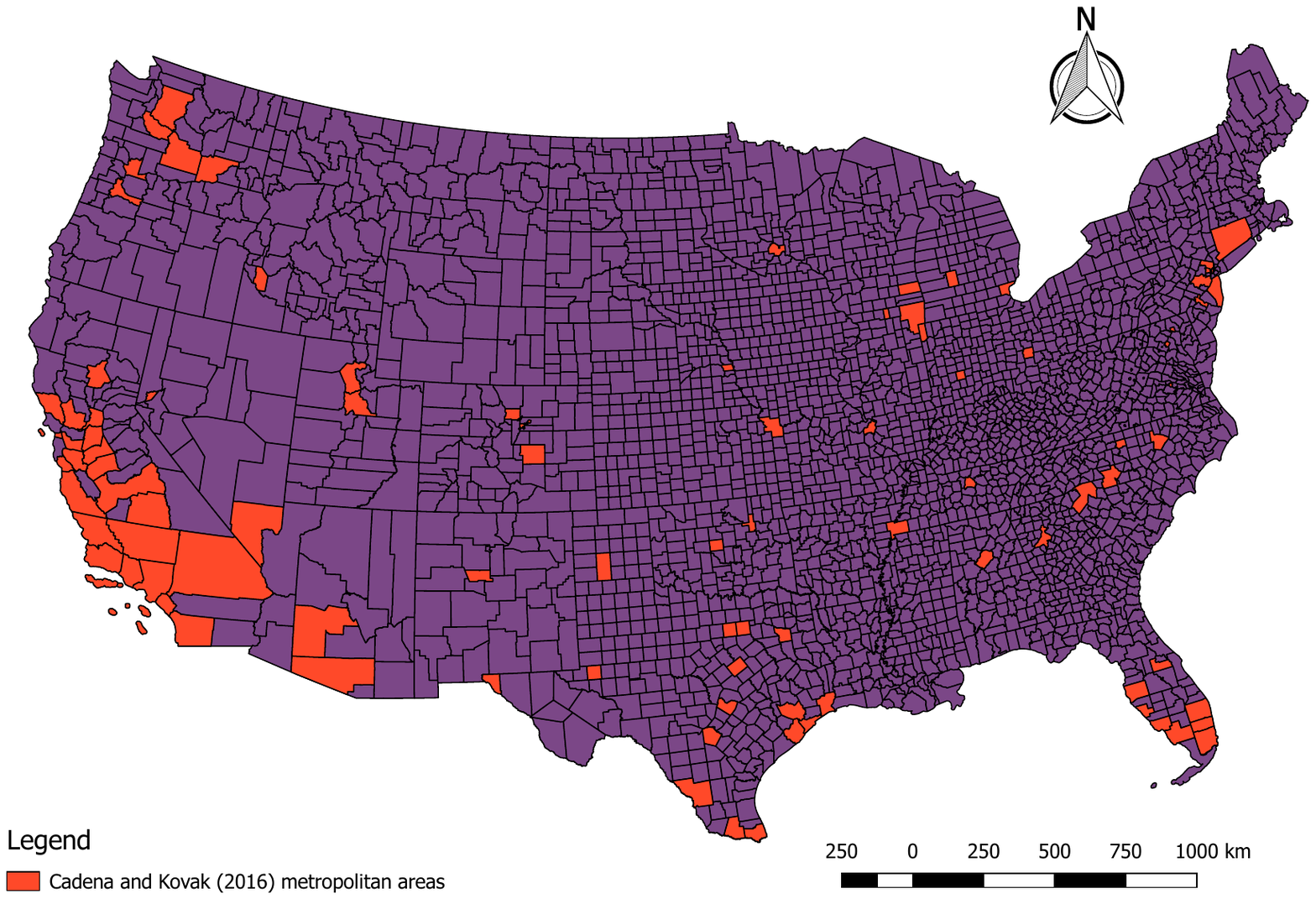}
\centering
\end{figure}

A potential issue with the estimation of \eqref{ck16_spec} is the potential existence of spatial dependency between the MSAs used in the analysis. Figure \ref{metck16} shows the 95 MSAs included in the \citet{ck16} IV analysis, revealing clear spatial heterogeneity. In order to account for potential spatial correlation we construct a SWM using a binary distance-based cut-off, where $w_{ij}=1$ if the distance between the metropolitan areas is less than $A$ kilometres and zero otherwise. We consider cut-off distances ($A$) of 500km, 600km, and 700km. Note 500km is the smallest distance that ensures every metropolitan area has at least one neighbour.

\begin{table}[t]
  \centering
  \caption{standardised Moran's $I$ of first and second stage \citep{ck16}}
  \label{tab:Z12ck}
  \begin{threeparttable}
    {\footnotesize
    \begin{tabular}{lccccc}
      \hline
      \T SWM &&&&& Other \\
      \B cut-off & All & Native-born & Foreign-born & Mexican-born & foreign-born \\
      \cline{2-6}
      \multicolumn{6}{l}{\qquad \emph{Panel A: Men, high school or less} \T\B }   \\
      \T  500km &  10.1$^{***}$, 3.25$^{***}$ & 11.36$^{***}$, 1.04 & 8.83$^{***}$, 2.86$^{***}$ & 9.7$^{***}$, 2.53$^{**}$ & 9.79$^{***}$, 2.7$^{***}$ \\
          600km & 10.34$^{***}$, 4.71$^{***}$ & 11.71$^{***}$, 2.34$^{**}$ & 8.65$^{***}$, 3.54$^{***}$ & 9.17$^{***}$, 2.22$^{**}$ & 10.02$^{***}$, 2.7$^{***}$ \\
      \B  700km & 10.59$^{***}$, 4.33$^{***}$ & 11.96$^{***}$, 2.86$^{***}$ & 8.83$^{***}$, 3.88$^{***}$ & 9.38$^{***}$, 2.6$^{***}$ & 10.25$^{***}$, 2.28$^{**}$ \\
      \multicolumn{6}{l}{\qquad \emph{Panel B: Men, some college or more} \T\B }   \\
      \T  500km & 10.43$^{***}$, 0.37 & 10.67$^{***}$, 0.01 & 9.93$^{***}$, 2.91$^{***}$ & 12.05$^{***}$, 0.25 & 10.35$^{***}$, 1.89$^{*}$ \\
          600km & 11.19$^{***}$, 1.21 & 11.58$^{***}$, 0.36 & 10.14$^{***}$, 2.53$^{**}$ & 11.76$^{***}$, -0.08 & 11.19$^{***}$, 1.84$^{*}$ \\
      \B  700km & 11.31$^{***}$, 1.39 & 11.8$^{***}$, 0.22 & 10$^{***}$, 2.01$^{**}$ & 11.84$^{***}$, 0.35 & 11.17$^{***}$, 1.3 \\
      \multicolumn{6}{l}{\qquad \emph{Panel C: Women, high school or less} \T\B }   \\
      \T  500km & 8.35$^{***}$,  2.47$^{**}$ & 11.41$^{***}$, 1.71$^{*}$ & 4.49$^{***}$, 3.03$^{***}$ & 4.21$^{***}$, 3.39$^{***}$ & 8.79$^{***}$, 1.53 \\
          600km &    9$^{***}$,  2.96$^{***}$ & 12.31$^{***}$, 2.43$^{**}$ & 4.19$^{***}$, 4.41$^{***}$ & 3.81$^{***}$, 3.44$^{***}$ & 8.82$^{***}$, 1.53 \\
      \B  700km & 9.34$^{***}$,  2.85$^{***}$ & 12.84$^{***}$, 2.3$^{**}$ & 4.13$^{***}$, 5.01$^{***}$ & 3.97$^{***}$, 4.04$^{***}$ & 8.98$^{***}$, 1.36 \\
      \multicolumn{6}{l}{\qquad \emph{Panel D: Women, some college or more} \T\B }   \\
      \T  500km &  9.14$^{***}$, 2.77$^{***}$ & 8.87$^{***}$, 1.28 & 9.89$^{***}$, 2.92$^{***}$ & 6.53$^{***}$, 0.1 & 10.5$^{***}$, 3.15$^{***}$ \\
          600km & 10.41$^{***}$, 3.06$^{***}$ & 10.12$^{***}$, 1.64 & 11.14$^{***}$, 3.87$^{***}$ & 6.72$^{***}$, 0.14 & 11.9$^{***}$, 3.91$^{***}$ \\
      \B  700km & 11.27$^{***}$, 3.53$^{***}$ & 10.98$^{***}$, 2.03$^{**}$ & 11.71$^{***}$, 2.96$^{***}$ & 6.83$^{***}$, -0.19 & 12.53$^{***}$, 3.4$^{***}$ \\
      \hline
     \end{tabular}
     }
  {\footnotesize
  \begin{tablenotes}
      \item \emph{Note}: first stage, second stage. $^{*}$p$<$0.1; $^{**}$p$<$0.05; $^{***}$p$<$0.01.
  \end{tablenotes}
  }
  \end{threeparttable}
\end{table}

Table \ref{tab:Z12ck} shows the standardised Moran's $i$ for the first and second stages of \citet{ck16} IV regressions obtained for the three SWMs considered. This exercise shows that the standardised Moran's $i$ of the first stage is always significant at the one percent level, and the second stage is also significant at the ten percent level in most configurations. In almost all cases, the first-stage has a substantially higher level of spatial correlation than the second-stage. For low-educated Mexican-born male migrants, the standardised Moran's $i$ is significant at the five percent level in both stages for all three SWMs, with a test statistic three times larger in the first than second stage. Given the presence of spatial dependence in the data there is a legitimate question as to how the IV estimates in \eqref{ck16_spec} might be affected. As explained in the introduction, this setting provides a realistic use-case for Mi-SL: the research question focuses on the impact of an endogenous covariate (the movements of Mexican-born lower skill workers) in a context where the presence of spatial dependence between the observational units (MSAs) potentially invalidates IV estimation. The spatial process present in the data can be detected in a straightforward manner, however accurately specifying it would go beyond the scope of the research question, and thus the researcher might well prefer to simply control for it, as a nuisance parameter.

\begin{table}[t]
  \centering
  \caption{Mi-2SL results of \citet{ck16} with 500km SWM cut-off}
  \label{ckMi-2SLl_500}
  \begin{threeparttable}
    {\footnotesize
    \begin{tabular}{lccccc}
      \hline
      \T  &&&&& Other \\
      \B  & All & Native-born & Foreign-born & Mexican-born & foreign-born \\
      \cline{2-6}
      \multicolumn{6}{l}{\qquad \emph{Panel A: Men, high school or less} \T\B }   \\
  \T  Change in log of group-specific &  0.274 &  0.128 &  0.327 & 1.262 & -0.597 \\
      employment                      & (0.182) & (0.092) & (0.438) & (0.359) & (0.32) \\
      First stage F-statistic         & 62.74 & 38.09 & 71.53 & 101.39 & 724.7 \\
      Partial F-statistic (Bartik)    & 60.89 & 43.37 & 37.09 & 55.84 & 112.96 \\
  \B  Number of vecs [1st,2nd]        & 7[7,0] & 11[11,0] & 5[5,0] & 13[13,0] & 3[3,0] \\
      \multicolumn{6}{l}{\qquad \emph{Panel B: Men, some college or more} \T\B }   \\
  \T  Change in log of group-specific &  0.203 &  0.304 & -0.233 &  0.307 & -0.035 \\
      employment                      & (0.172) & (0.241) & (0.191) & (0.723) & (0.252) \\
      First stage F-statistic         & 28.32 & 23.04 & 29.7 & 136.59 & 22.02 \\
      Partial F-statistic (Bartik)    & 50.3 & 55.95 & 44.61 & 110.85 & 86.63 \\
  \B  Number of vecs [1st,2nd]        & 6[6,0] & 8[8,0] & 3[3,0] & 19[19,0] & 6[6,0] \\
      \multicolumn{6}{l}{\qquad \emph{Panel C: Women, high school or less} \T\B }   \\
  \T  Change in log of group-specific &  0.148 & -0.389 &  0.272 & 1.811 & -0.737 \\
      employment                      & (0.158) & (0.237) & (0.504) & (0.665) & (0.458) \\
      First stage F-statistic         & 41.51 & 34.32 & 15.21 & 6.04 & 49.4 \\
      Partial F-statistic (Bartik)    & 64.18 & 57.57 & 26.76 & 13.74 & 95.59 \\
  \B  Number of vecs [1st,2nd]        & 1[1,0] & 6[6,0] & 0[0,0] & 0[0,0] & 1[1,0] \\
      \multicolumn{6}{l}{\qquad \emph{Panel D: Women, some college or more} \T\B }   \\
      Change in log of group-specific & -0.02 & -0.037 & -0.581 &  0.061 & -0.987 \\
      employment                      & (0.301) & (0.381) & (0.517) & (1.118) & (0.508) \\
      First stage F-statistic         & 12.32 & 10.64 & 22.37 & 24.51 & 21.14 \\
      Partial F-statistic (Bartik)    & 9.62 & 8.17 & 26.81 & 36.29 & 24.32 \\
  \B  Number of vecs [1st,2nd]        & 1[1,0] & 6[6,0] & 0[0,0] & 0[0,0] & 1[1,0] \\
      \hline
     \end{tabular}
     }
  {\footnotesize
  \begin{tablenotes}
      \item \emph{Note}: First stage fitted values from Lasso estimates (step 3 in Algorithm \ref{Mi-2SLalg}) and the cut-off for the SWM is 500km. Robust standard errors in parentheses.
  \end{tablenotes}
  }
  \end{threeparttable}
\end{table}


Table \ref{ckMi-2SLl_500} shows the estimates obtained using Mi-2SLl (the first stage fitted values from the Lasso estimates, step 3 in Algorithm \ref{Mi-2SLalg}) with the 500km cut-off SWM. No eigenvectors are selected in any of the second stages, due to the lower levels of spatial correlation in each of the second stages, so the fitted values from Lasso and post-Lasso yield the same results. Tables of results obtained with the larger the larger cut-off SWMs, which were run as a robustness check, are provided in appendix \ref{extra_tabs}. The Mi-2SL results do not change the qualitative conclusion of \citet{ck16} that low-educated Mexican-born migrants respond positively to changes in employment. However, for low-educated Mexican-born males we find the magnitude of the coefficient increases by approximately a standard error. More generally, a key general impact of the included eigenvectors is a substantial improvement in the first stage F-statistic and partial F-statistic. For example, for low-educated Mexican-born males, the first-stage F-statistic and partial F-statistic increase from 8.28 and 11.94 to 101.39.35 and 55.84 respectively. This improvement in the first-stage estimates leads to an increase in the precision of the predicted values and this of the second-stage estimates, which can be seen by the reduction in the estimated standard errors on employment change from 0.468 to 0.359. The fact that the partial F-statistic is now smaller that the full F-statistic also implies that the Bartik is now having a stronger positive effect, at least in the case of low-educated Mexican-born migrants.

\section{Conclusion}\label{conc}

In conclusion, we have proposed a new two-stage lasso-based procedure, called Moran's $I$ 2 stage Lasso (Mi-2SL), to estimate classical regression parameters of endogenous variables in the presence of spatial correlation of an unknown functional form. Under the assumption that the relevant set of eigenvectors is known, that an appropriate mixing condition holds, that some restriction exists on the spatial structure, and some assuming some additional regularity conditions, we show that the Mi-2SLparameter estimates are consistent and asymptotic normal.

Our simulations results establish that the Mi-2SL estimators offer good performance in small samples against a range of competing estimator in the presence of spatial correlation, both in terms of bias and asymptotic variance. In particular, performance is equivalent to IV estimation when spatial correlation is absent or its impact is small: in such cases the Lasso procedure simply fails to select any eigenvectors and the resulting estimator boils down to a simple 2SLS. Should a researcher need to estimate an IV specification but then detect the presence of spatial dependence with a given SWM, our recommendation is therefore to instead run Mi-2SL as a protection against the adverse effect of that dependence on the estimates. At worst, if the spatial dependence is weak, the two estimators will produce the same estimates and thus Mi-2SL does no harm. At best, Mi-2SL will effectively control for the spatial dependence. Our empirical application, where we replicate the IV results of \citet{ck16}, demonstrates the benefits of using Mi-2SL in the presence of clear spatial dependence, by improving the first-stage partial F-statistic and full F-statistic, and reducing the second stage standard errors.

Several avenues of further research involve investigating the robustness of Mi-2SL to various misspecifications. The first is the fact that the set of relevant eigenvectors $\Omega$ need to be estimated, and therefore the robustness of consistency and asymptotic normality in the presence of mistakes in eigenvector selection should be investigated. A related direction is robustness to the specification of the SWM. The results obtained here use the true SWM from the data-generating process, however in empirical settings, the true SWM is unobserved and it is likely that the empirical SWM will be misspecified in some way. Clearly, if the empirical SWM is correlated enough to the truth the Moran's $I$ test will have power to detect the correlation. However the methodology would benefit from a greater theoretical understanding of how performance will degrade as the misspecification of the empirical SWM increases.

\onehalfspacing 

\bibliographystyle{agsm}
\bibliography{reference}

\doublespacing 

\begin{appendices}
\section{Additional Lemmas and proofs main results}\label{app:proofs2}

The following Lemma shows that as linear transformation (denoted $h_c$) is a Lipschitz function, then taking $f_c=f(h_c(\cdot))$ is also a Lipschitz function, as long as the linear transformation is bounded.

\begin{lem}\label{lintran}
  If $f(r_i)$ is a bounded real Lipschitz function with $f\in\pazocal{L}_{a}$ and $\{r_i\in\R\}$ are triangular arrays and $h_c(r_i)$ is a bounded real linear Lipschitz function which linearly transforms $r_i$ by the triangular arrays $\{c_i\}$, $c_i\in\R$ with $\max_i|c_i|<\infty$. Then $f_c(r_i)=f\big(h_c(r_i)\big)$ is also a bounded real Lipschitz with $f_c\in f$.
\end{lem}

\begin{proof}[Proof of Lemma~\ref{lintran}]
  Let the linear function $h_c$ be multiplicative i.e. $h_c(r_i):=c_ir_i$, $\R\to \R$. $h_c$  is Lipschitz iff there exists a constant $K_h$ such that $|h_c(x_1)- h_c(x_2)|=|c_ix_1- c_ix_2|  \leq K_h |(x_1-x_2)|$ where $x_1$ and $x_2$ are points in $\R$. Note by the Cauchy-Schwartz inequality $(c_i(x_1-x_2))^2 \leq c_i^2(x_1-x_2)^2$. Thus, $h_c(\cdot)$ is Lipschitz as $K_h=|c_i|<\infty$.

  Now as $f$ is a bounded real Lipschitz function, $|f(c_ix_1)- f(c_ix_2)| \leq K_f |c_i'x_1-c_i'x_2|$ thus:
  \begin{align}
    |f_c(x_1)-f_c(x_1) | \enspace = \enspace |f(c_ix_1) -f(c_ix_2) | & \enspace \leq \enspace  K_f |c_i(x_1-x_2)| \\
    & \enspace  \leq \enspace  K_c |(x_1-x_2)|\label{cs1}
  \end{align}
  where \eqref{cs1} uses the Cauchy-Schwartz inequality and $K_c:= K_fK_h$.

  Similarly let the linear function $h_c$ be additive i.e. $h_c(r_i):=c_i+r_i$, $\R\to \R$. $h_c$ is Lipschitz iff there exists a constant $K_h$ such that $|h_c(x_1)- h_c(x_2)|=|(c_i+x_1)- (c_i+x_2)|  \leq  |(x_1-x_2)|$, note $K_h=1$. Thus,
  \begin{align*}
    |f_c(x_1)-f_c(x_1) | \enspace = \enspace |f(c_i+x_1) -f(c_i+x_2) | & \enspace \leq \enspace K_f |(c_i+x_1)-(c_i+x_2)| = K_f |x_1-x_2|\\
    & \enspace = \enspace K_c |(x_1-x_2)|
  \end{align*}
  where $K_f=K_cK_h=K_c$.
\end{proof}

\begin{proof}[Proof of Lemma~\ref{eigtran}]
  As $e_{jk}$ and $\lambda_k$ satisfy the linear transformation requirements in Lemma \ref{lintran}, these transformations are $f_c\in f$ and $g_c\in g$, so we have:
  \begin{align*}
    \left|\cov\left(f\left(\sum_{j\in N} w_{a}e_{jk}/\lambda_k\right),g\left(\sum_{j\in N} w_{b}e_{jk}/\lambda_k\right)|\C\right)\right|& \enspace = \enspace \left|\cov(f_c(w_{a}),g_c(w_{b})|\C)\right| \\
    & \enspace \leq \enspace  \psi_{a,b}(f_c,g_c)\mu_{r} \quad \as
  \end{align*}
  Where the inequality holds due to \eqref{covmix}.
\end{proof}

\begin{lem}\label{lintran2}
   Suppose the triangular arrays $\{a_{ki}\}$, $a_{ki}\in\R$ satisfy Assumption \ref{as:lln2}.1 with dependence coefficient $\{\mu_r\}$ and the columns of $\bm{A}$ are orthonormal ($||\bm{a}_i||_2^2=1 \; \forall i$). Then the array $\{\sum^n_{k=1}a_{ki}a_{kj}\},\; i=1,\ldots,s, \; j=1,\ldots,s$ is conditionally $\psi$-dependent given $\{\C\}$ with the dependence coefficients $\{\mu_r\}$.
   \begin{equation*}
     \left|\cov\left(f\left(\sum_{k\in N} a_{a}a_{kj}\right),g \left(\sum_{k\in N}a_{b}a_{kj}\right)|\C\right)\right| \enspace \leq \enspace \psi_{a,b}(f_c,g_c)\mu_{r} \quad \as
   \end{equation*}
   where $a_{a}=a_{ka}$ and $a_{b}=a_{kb}$.
\end{lem}

\begin{proof}[Proof of Lemma~\ref{lintran2}]
  As $||\bm{a}_j||_2^2=1$ satisfy the linear transformation requirements in Lemma \ref{lintran}, these transformations are $f_c\in f$ and $g_c\in g$, thus we have:
  \begin{equation*}
  \left|\cov\left(f\left(\sum_{k\in N} a_{a}a_{kj}\right),g \left(\sum_{k\in N}a_{b}a_{kj}\right)|\C\right)\right| = \left|\cov(f_c(a_{a}),g_c(a_{b})|\C)\right|\leq \psi_{a,b}(f_c,g_c)\mu_{r}\quad \as
 \end{equation*}
  where the inequality holds due to \eqref{covmix}.
\end{proof}

\begin{lem}\label{lintran3}
   Suppose the triangular array $\{a_{ki}\}$, $a_{ki}\in\R$ satisfy Assumption \ref{as:lln2}.1 with dependence coefficient $\{\mu_r\}$. $\forall n \geq 1$ let $\{b_{kj}\}$, $b_{kj}\in\R$ be a sequence of $\C$-measurable random variables with $\sup_{n\geq 1} \max_{i\in N}\big(\E[|b_{kj}|^{2}|\C]\big)^{1/2}<\infty, \;\forall j$. Then the
  array $\{\sum^n_{k=1}a_{ki}b_{kj}\}$, $i=1,\ldots,s$, $j=1,\ldots,d$ is conditionally $\psi$-dependent given $\{\C\}$, with dependence coefficients $\{\mu_r\}$.
  \begin{equation*}
    \left|\cov\left(f\left(\sum_{k\in N} a_{a}b_{kj}\right),g \left(\sum_{k\in N}a_{b}b_{kj}\right)|\C\right)\right| \enspace \leq \enspace \psi_{a,b}(f_c,g_c)\mu_{r} \quad \as
  \end{equation*}
   where $a_{a}=a_{ka}$ and $a_{b}=a_{kb}$.
\end{lem}

\begin{proof}[Proof of Lemma~\ref{lintran3}]
  As $\sup_{n\geq 1} \max_{i\in N}\big(\E[|b_{kj}|^{2}|\C]\big)^{1/2}<\infty, \;\forall j$ satisfy the linear transformation requirements in Lemma \ref{lintran}, these transformations are $f_c\in f$ and $g_c\in g$, thus:
  \begin{equation*}
   \left|\cov\left(f\left(\sum_{k\in N} a_{a}b_{kj}\right),g \left(\sum_{k\in N}a_{b}b_{kj}\right)|\C\right)\right| = \left|\cov(f_c(a_{a}),g_c(a_{b})|\C)\right|\leq \psi_{a,b}(f_c,g_c)\mu_{r} \quad \as
  \end{equation*}
  where the inequality holds due to \eqref{covmix}.
\end{proof}

Note, a special case of Lemma \ref{lintran3} is the case $d=1$, i.e. $\bm{B}$ is a column vector.

\begin{proof}[Proof of Theorem~\ref{theo:IVsfplim}]
  Starting from the 2SLS solutions of \eqref{yesfsupp}:
  \begin{equation*}
    \hat{\bm{\Upsilon}}_{\Omega} = \big((\bm{G}_{\Omega}'\bm{Z}_{\Omega}/n)(\bm{Z}_{\Omega}'\bm{Z}_{\Omega}/n)^{-1}(\bm{Z}_{\Omega}'\bm{G}_{\Omega}/n)\big)^{-1}(\bm{G}_{\Omega}'\bm{Z}_{\Omega}/n)(\bm{Z}_{\Omega}'\bm{Z}_{\Omega}/n)^{-1}(\bm{Z}_{\Omega}'\bm{y}/n)
  \end{equation*}
  Substituting in \eqref{yesfsupp} yields:
  \begin{align}
     \hat{\bm{\Upsilon}}_{\Omega} & \enspace = \enspace \big((\bm{G}_{\Omega}'\bm{Z}_{\Omega}/n)(\bm{Z}_{\Omega}'\bm{Z}_{\Omega}/n)^{-1}(\bm{Z}_{\Omega}'\bm{G}_{\Omega}/n)\big)^{-1} \\
    & \;\;\;\; \times    (\bm{G}_{\Omega}'\bm{Z}_{\Omega}/n)(\bm{Z}_{\Omega}'\bm{Z}_{\Omega}/n)^{-1}(\bm{Z}_{\Omega}'(\bm{G}_{\Omega} \bm{\Upsilon}_\Omega + \bm{\varepsilon})/n) \nonumber \\
     \hat{\bm{\Upsilon}}_{\Omega} - \bm{\Upsilon}_{\Omega} & \enspace = \enspace \big((\bm{G}_{\Omega}'\bm{Z}_{\Omega}/n)(\bm{Z}_{\Omega}'\bm{Z}_{\Omega}/n)^{-1}(\bm{Z}_{\Omega}'\bm{G}_{\Omega}/n)\big)^{-1} \\
      & \;\;\;\; \times (\bm{G}_{\Omega}'\bm{Z}_{\Omega}/n)(\bm{Z}_{\Omega}'\bm{Z}_{\Omega}/n)^{-1}(\bm{Z}_{\Omega}'\bm{\varepsilon}/n) \label{hu-u}
   \end{align}
Expressing $\bm{Z}_{\Omega}'\bm{Z}_{\Omega}/n$, $ \bm{G}_{\Omega}'\bm{Z}_{\Omega}/n$, $ \bm{Z}_{\Omega}'\bm{G}_{\Omega}/n$ and $\bm{Z}_{\Omega}'\bm{\varepsilon}/n$ in block-wise form yields:
   \begin{align*}
   \bm{Z}_{\Omega}'\bm{Z}_{\Omega}/n & = \begin{bmatrix} \bm{X}_1'\bm{X}_1/n & \bm{X}_1'\bm{Z}_2/n & \bm{X}_1'\bm{E}_{\Omega}/n \\
     \bm{Z}_2'\bm{X}_1/n & \bm{Z}_2'\bm{Z}_2/n & \bm{Z}_2'\bm{E}_{\Omega}/n \\
   \bm{E}_{\Omega}'\bm{X}_1/n & \bm{E}_{\Omega}'\bm{Z}_2/n & \bm{E}_{\Omega}'\bm{E}_{\Omega}/n  \end{bmatrix}\\
   \bm{G}_{\Omega}'\bm{Z}_{\Omega}/n & = \begin{bmatrix} \bm{X}_1'\bm{X}_1/n & \bm{X}_1'\bm{Z}_2/n & \bm{X}_1'\bm{E}_{\Omega}/n \\
   \bm{x}_2'\bm{X}_1/n & \bm{x}_2'\bm{Z}_2/n & \bm{x}_2'\bm{E}_{\Omega}/n \\
   \bm{E}_{\Omega}'\bm{X}_1/n & \bm{E}_{\Omega}'\bm{Z}_2/n & \bm{E}_{\Omega}'\bm{E}_{\Omega}/n  \end{bmatrix} \\
   \bm{Z}_{\Omega}'\bm{G}_{\Omega}/n & = \begin{bmatrix} \bm{X}_1'\bm{X}_1/n & \bm{X}_1'\bm{x}_2/n & \bm{X}_1'\bm{E}_{\Omega}/n \\
   \bm{Z}_2'\bm{X}_1/n & \bm{Z}_2'\bm{x}_2/n & \bm{Z}_2'\bm{E}_{\Omega}/n \\
  \bm{E}_{\Omega}'\bm{X}_1/n & \bm{E}_{\Omega}'\bm{x}_2/n & \bm{E}_{\Omega}'\bm{E}_{\Omega}/n  \end{bmatrix}\\
   \bm{Z}_{\Omega}'\bm{\varepsilon}/n & = \begin{bmatrix} \bm{X}_1'\bm{\varepsilon}/n  \\
   \bm{Z}_2'\bm{\varepsilon}/n  \\
   \bm{E}_{\Omega}'\bm{\varepsilon}/n   \end{bmatrix}
   \end{align*}
The elements of $\bm{X}_1$, $\bm{x}_2$ and $\bm{Z}_2$ are triangular arrays of real number that are bounded in absolute value (Assumptions \ref{as:spatial2}.2). Additionally by Assumptions \ref{as:spatial2}.3 and the LLN of triangular arrays, we have the following block-wise convergences in probability as $n \to \infty$:
\begin{equationarray*}{l@{\qquad}l@{\qquad}l}
  \bm{X}_1'\bm{X}_1/n \to_p \E[\bm{X}_{11}'\bm{X}_{11}] & \bm{X}_1'\bm{Z}_2/n \to_p \E[\bm{X}_{11}'\bm{Z}_{21}] & \bm{Z}_2'\bm{X}_1/n \to_p \E[\bm{Z}_2'\bm{X}_1] \\
  \bm{Z}_2'\bm{Z}_2/n \to_p \E[\bm{Z}_{21}'\bm{Z}_{21}] & \bm{x}_2'\bm{X}_1/n \to_p \E[\bm{x}_{21}'\bm{X}_{21}] & \bm{X}_1'\bm{x}_2/n \to_p \E[\bm{X}_{11}'\bm{x}_{21}] \\
  \bm{Z}_2'\bm{x}_2/n \to_p \E[\bm{Z}_{21}'\bm{x}_{21}] & \bm{x}_2'\bm{Z}_2/n \to_p \E[\bm{x}_{21}'\bm{Z}_{21}] & \bm{X}_1'\bm{\varepsilon}/n \to_p \E[\bm{X}_{11}'\bm{\varepsilon}]=0 \\
  \multicolumn{3}{l}{\bm{Z}_2'\bm{\varepsilon}/n \to_p \E[\bm{Z}_{21}'\bm{\varepsilon_{\pi1}}]=0 +o_p(1)}
\end{equationarray*}
By Lemma \ref{eigtran} - \ref{lintran3} terms involving $\bm{E}_{\Omega}$ are weakly $\psi$-dependent with dependence coefficient $\{\mu_r\}$, thus, under Assumptions \ref{as:lln2} and the LLN of \citet{LT_KMS20} we have the following block-wise almost sure convergences as $n \to \infty$:
\begin{equationarray*}{l@{\quad}l@{\quad}l}
  \bm{X}_1'\bm{E}_{\Omega}/n \to_{a.s.} \E[\bm{X}_{11}'\bm{E}_{\Omega 1}|\C] &\bm{Z}_{2}'\bm{E}_{\Omega}/n \to_{a.s.} \E[\bm{Z}_{21}'\bm{E}_{\Omega1}|\C] & \bm{E}_{\Omega}'\bm{X}_1/n \to_{a.s.} \E[\bm{E}_{\Omega 1}'\bm{X}_{11}|\C]\\ \bm{E}_{\Omega}'\bm{Z}_2/n \to_{a.s.} \E[\bm{E}_{\Omega 1}'\bm{Z}_{21}|\C] & \bm{x}_2'\bm{E}_{\Omega}/n \to_{a.s.} \E[\bm{x}_{21}'\bm{E}_{\Omega 1}|\C] & \bm{E}_{\Omega}'\bm{x}_2/n \to_{a.s.} \E[\bm{E}_{\Omega 1}'\bm{x}_{21}|\C]\\
  \bm{E}_{\Omega}'\bm{E}_{\Omega}/n \to_{a.s.} \E[\bm{E}_{\Omega 1}'\bm{E}_{\Omega 1}|\C] & \multicolumn{2}{l}{\bm{E}_{\Omega}'\bm{\varepsilon}/n \to_{a.s.} \E[\bm{E}_{\Omega 1}'\bm{\varepsilon_{\pi 1}}|\C]=0+o_p(1)}
\end{equationarray*}
Combining the block elements together, we have $\bm{Z}_{\Omega}'\bm{Z}_{\Omega}/n \to_p \E[\bm{Z}_{\Omega 1}'\bm{Z}_{\Omega 1}|\C]$, $\bm{G}_{\Omega}'\bm{Z}_{\Omega}/n \to_p \E[\bm{G}_{\Omega 1}'\bm{Z}_{\Omega 1}|\C]$, $\bm{Z}_{\Omega}'\bm{G}_{\Omega}/n \to_p \E[\bm{Z}_{\Omega 1}'\bm{G}_{\Omega 1}|\C]$ and $\bm{Z}_{\Omega}'\bm{\varepsilon}/n \to_p \E[ \bm{Z}_{\Omega 1}'\bm{\varepsilon}_{\pi 1}|\C]=0$. Applying the Continuous Mapping Theorem we have:
\[
   \begin{aligned}
     \hat{\bm{\Upsilon}}_{\Omega} - \bm{\Upsilon}_{\Omega} \enspace & \to_p \enspace  & \big(\E[\bm{G}_{\Omega 1}'\bm{Z}_{\Omega 1}|\C]\E[\bm{Z}_{\Omega 1}'\bm{Z}_{\Omega 1}|\C]^{-1}
     \E[\bm{Z}_{\Omega 1}'\bm{G}_{\Omega 1}|\C]\big)^{-1} \\
      & &  \times \E[\bm{G}_{\Omega 1}'\bm{Z}_{\Omega 1}|\C]\E[\bm{Z}_{\Omega 1}'\bm{Z}_{\Omega 1}|\C]^{-1}\E[\bm{Z}_{\Omega 1}'\bm{\varepsilon}_{\pi 1}|\C] \\
      & = 0 &
   \end{aligned}
\]
\end{proof}
\begin{proof}[Proof of Theorem~\ref{theo:IVsfdist}]
  Multiplying \eqref{hu-u} by $\sqrt{n}$ gives:
  \begin{align*}
    \sqrt{n}(\hat{\bm{\Upsilon}}_{\Omega} - \bm{\Upsilon}_{\Omega}) \enspace = \enspace & \big((\bm{G}_{\Omega}'\bm{Z}_{\Omega}/n)(\bm{Z}_{\Omega}'\bm{Z}_{\Omega}/n)^{-1}(\bm{Z}_{\Omega}'\bm{G}_{\Omega}/n)\big)^{-1} \\
     & \times (\bm{G}_{\Omega}'\bm{Z}_{\Omega}/n)(\bm{Z}_{\Omega}'\bm{Z}_{\Omega}/n)^{-1}(\bm{Z}_{\Omega}'\bm{\varepsilon}/\sqrt{n})
  \end{align*}
  Under Assumptions \ref{as:spatial2}-\ref{as:lln2}, the LLN of triangular arrays and the LLN of \citet{LT_KMS20} that matrices involving $\bm{G}_{\Omega}$ and $\bm{Z}_{\Omega}$ are $O_p(1)$ (proved in Theorem \ref{theo:IVsfplim}). We now need to look at the behaviour of:
  \begin{equation*}
    \bm{Z}_{\Omega}'\bm{\varepsilon}/\sqrt{n} = \begin{bmatrix} \bm{X}_1'\bm{\varepsilon}/\sqrt{n}  \\
    \bm{Z}_2'\bm{\varepsilon}/\sqrt{n}  \\
    \bm{E}_{\Omega}'\bm{\varepsilon}/\sqrt{n}   \end{bmatrix}
  \end{equation*}

  Given $\varepsilon_i$ is a triangular array of identically distributed random variables that is (jointly) independently distributed for each $n$ with $\E[\varepsilon_i]=0$ and $\E[\varepsilon_i^2]=\sigma_\varepsilon^2 < \infty$ (Assumption \ref{as:spatial2}.2), and
  $\E[\bm{X}_{11}'\bm{X}_{11}|\C]$ and $\E[\bm{Z}_{21}'\bm{Z}_{21}|\C]$ are finite and non-singular (implied by Assumption \ref{as:spatial2}.2 and \ref{as:lln2}.2). Then the central limit theorem for triangular arrays implies $\bm{X}_{11}'\bm{\varepsilon}_{\pi}/\sqrt{n} \enspace \to_d \enspace N(0, \sigma_\varepsilon^2 plim_{n \to \infty}[\bm{X}_1'\bm{X}_1/n|\C] )+ O_p(1/n)$
   and $\bm{Z}_2'\bm{\varepsilon_{\pi}}/\sqrt{n} \enspace \to_d \enspace N(0, \sigma_\varepsilon^2plim_{n \to \infty}[\bm{Z}_2'\bm{Z}_2/n|\C] )$ $+O_p(1/n)$.

  Lemma \ref{lintran}, \ref{eigtran} and \ref{lintran3} insure that the element of $\bm{E}_{\Omega}'\bm{\varepsilon}$ are $\psi$-dependent with dependence coefficients $\mu_r$.
  We can show  that $\bm{Z}_{\Omega}'\bm{\varepsilon}$ has a finite second moment. By Minkowski's inequality we have:
   \begin{align*}
        \left(\E[|\varepsilon_i|^4|\C]\right)^{1/4} &\enspace = \enspace \left(\E\left[|y_i-\sum^{(k_1+1+s)}_{j=1}g_{ij,\Omega}\Upsilon_{j,\Omega}|^4|\C\right]\right)^{1/4}  \\
        &\enspace \leq \enspace \left(\E\left[|y_i|^4|\C\right]\right)^{1/4}+\left(\E\left[\sum^{(k_1+1+s)}_{j=1}|g_{ij,\Omega}|^4|\C\right]\right)^{1/4}\sum^{(k_1+1+s)}_{j=1}\Upsilon_{j,\Omega} \enspace < \enspace \infty
   \end{align*}
  under Assumption \ref{as:clt2}.1, for $i=1,\ldots ,n$. Then by the Cauchy-Schwarz inequality:
  \begin{equation*}
    \E\left[\sum^{(k_1+1+s)}_{j=1}|z_{ij,\Omega}\varepsilon_i|^2|\C\right] \enspace \leq \enspace  \E\left[\sum^{(k_1+1+s)}_{j=1}|z_{ij,\Omega}|^4|\C\right]\E\left[|\varepsilon_i|^4|\C\right] \enspace  < \enspace \infty
  \end{equation*}
  under Assumption \ref{as:clt2}.1,  for $i=1,\ldots ,n$.
 Additionally given assumptions  \ref{as:spatial2}.3, \ref{as:lln2}.4 and \ref{as:clt2} , we can apply the CLT of \citet{LT_KMS20}, $\bm{E}_{\Omega}'\bm{\varepsilon}/\sqrt{n} \; \to_d \; N(0, \sigma_\varepsilon^2plim_{n \to \infty}[\bm{E}_{\Omega}'\bm{E}_{\Omega}/n|\C])$. Thus, we have:
   \begin{align*}
     \sqrt{n}(\hat{\bm{\Upsilon}}_{\Omega} - \bm{\Upsilon}_{\Omega}) \enspace & = \enspace  \big((\bm{G}_{\Omega}'\bm{Z}_{\Omega}/n)(\bm{Z}_{\Omega}'\bm{Z}_{\Omega}/n)^{-1}(\bm{Z}_{\Omega}'\bm{G}_{\Omega}/n)\big)^{-1} \\
      & \qquad \times (\bm{G}_{\Omega}'\bm{Z}_{\Omega}/n)(\bm{Z}_{\Omega}'\bm{Z}_{\Omega}/n)^{-1}(\bm{Z}_{\Omega}'\bm{\varepsilon}/\sqrt{n}) \\
      \enspace & \to_d \enspace N\left(0, n \left(plim_{n \to \infty}\big([\bm{G}_{\Omega}'\bm{Z}_{\Omega}|\C][\bm{Z}_{\Omega}'\bm{Z}_{\Omega}|\C]^{-1}
      [\bm{Z}_{\Omega}'\bm{G}_{\Omega}|\C]\big)^{-1} \right)\sigma_\varepsilon^2\right)
   \end{align*}

\end{proof}

\newpage
\section{Additional simulation tables}\label{sims2full}

\setlength{\tabcolsep}{5pt} 

\begin{table}[!h]
  \centering
  \caption{Results for $n=250$ and $\omega=0.4$}
  \label{tab:n250,o4}
  \begin{threeparttable}
    {\footnotesize
      \begin{tabular}{ccclrcccrccc}
        \hline
        \multicolumn{3}{c}{Experiment} & \T \B & \multicolumn{4}{c}{Rewiring prob. $p=0.4$} & \multicolumn{4}{c}{Rewiring prob. $p=0.8$} \\
        \T \B $\rho$ & $\zeta_{3,1}$ & $\zeta_{3,2}$ & Estimator & bias & MSE & AASE & Vecs & bias & MSE & AASE  &  Vecs \\
        \hline
        \multirow{5}{*}{0.4} & \multirow{5}{*}{0.4} & \multirow{5}{*}{  0} & \T  SimpOLS &  0.496 &  0.248 &  0.036 & - &  0.491 &  0.242 &  0.036 & - \\
                             &                      &                      &      SimpIV &  0.025 &  0.005 &  0.068 & - &  0.017 &  0.005 &  0.068 & - \\
                             &                      &                      &    2SLS-SAR & -0.005 &  0.004 &  0.065 & - & -0.009 &  0.004 &  0.065 & - \\
                             &                      &                      &     Mi-2SLl &  0.005 &  0.008 &  0.047 & [5,88] 89 &  0.000 &  0.007 &  0.049 & [4,76] 76 \\
                             &                      &                      & \B Mi-2SLpl &  0.008 &  0.007 &  0.049 & [5,70] 75 &  0.004 &  0.007 &  0.051 & [4,61] 65 \\

        \multirow{5}{*}{0.4} & \multirow{5}{*}{0.4} & \multirow{5}{*}{0.4} & \T  SimpOLS &  0.493 &  0.244 &  0.036 & - &  0.487 &  0.239 &  0.036 & - \\
                             &                      &                      &      SimpIV &  0.026 &  0.005 &  0.067 & - &  0.018 &  0.005 &  0.067 & - \\
                             &                      &                      &    2SLS-SAR & -0.006 &  0.004 &  0.064 & - & -0.010 &  0.004 &  0.065 & - \\
                             &                      &                      &     Mi-2SLl &  0.009 &  0.008 &  0.046 & [8,92] 93 &  0.002 &  0.007 &  0.048 & [6,79] 80 \\
                             &                      &                      & \B Mi-2SLpl &  0.011 &  0.007 &  0.048 & [8,68] 75 &  0.003 &  0.006 &  0.050 & [6,60] 65 \\

        \multirow{5}{*}{0.4} & \multirow{5}{*}{0.8} & \multirow{5}{*}{  0} & \T  SimpOLS &  0.507 &  0.258 &  0.035 & - &  0.500 &  0.251 &  0.035 & - \\
                             &                      &                      &      SimpIV &  0.035 &  0.006 &  0.067 & - &  0.026 &  0.005 &  0.067 & - \\
                             &                      &                      &    2SLS-SAR & -0.005 &  0.004 &  0.065 & - & -0.009 &  0.004 &  0.066 & - \\
                             &                      &                      &     Mi-2SLl &  0.018 &  0.008 &  0.049 & [49,48] 87 &  0.015 &  0.008 &  0.049 & [37,50] 76 \\
                             &                      &                      & \B Mi-2SLpl &  0.007 &  0.008 &  0.054 & [49,19] 67 &  0.004 &  0.007 &  0.055 & [37,21] 57 \\

        \multirow{5}{*}{0.4} & \multirow{5}{*}{0.8} & \multirow{5}{*}{0.4} & \T  SimpOLS &  0.504 &  0.255 &  0.034 & - &  0.496 &  0.248 &  0.034 & - \\
                             &                      &                      &      SimpIV &  0.037 &  0.006 &  0.066 & - &  0.027 &  0.005 &  0.066 & - \\
                             &                      &                      &    2SLS-SAR & -0.005 &  0.004 &  0.064 & - & -0.009 &  0.004 &  0.065 & - \\
                             &                      &                      &     Mi-2SLl &  0.012 &  0.008 &  0.050 & [66,34] 91 &  0.011 &  0.007 &  0.050 & [47,40] 78 \\
                             &                      &                      & \B Mi-2SLpl &  0.005 &  0.008 &  0.054 & [66,12] 77 &  0.003 &  0.007 &  0.054 & [47,16] 63 \\

        \multirow{5}{*}{0.8} & \multirow{5}{*}{0.4} & \multirow{5}{*}{  0} & \T  SimpOLS &  0.552 &  0.307 &  0.042 & - &  0.539 &  0.293 &  0.041 & - \\
                             &                      &                      &      SimpIV &  0.055 &  0.009 &  0.075 & - &  0.043 &  0.007 &  0.074 & - \\
                             &                      &                      &    2SLS-SAR & -0.008 &  0.004 &  0.065 & - & -0.012 &  0.004 &  0.066 & - \\
                             &                      &                      &     Mi-2SLl & -0.002 &  0.010 &  0.042 & [5,168] 168 & -0.009 &  0.009 &  0.042 & [4,152] 152 \\
                             &                      &                      & \B Mi-2SLpl &  0.013 &  0.011 &  0.042 & [5,157] 159 &  0.004 &  0.010 &  0.042 & [4,142] 145 \\

        \multirow{5}{*}{0.8} & \multirow{5}{*}{0.4} & \multirow{5}{*}{0.4} & \T  SimpOLS &  0.552 &  0.307 &  0.041 & - &  0.537 &  0.291 &  0.041 & - \\
                             &                      &                      &      SimpIV &  0.057 &  0.009 &  0.075 & - &  0.044 &  0.007 &  0.074 & - \\
                             &                      &                      &    2SLS-SAR & -0.009 &  0.004 &  0.064 & - & -0.012 &  0.004 &  0.065 & - \\
                             &                      &                      &     Mi-2SLl &  0.006 &  0.010 &  0.041 & [8,172] 172 & -0.006 &  0.009 &  0.042 & [6,156] 156 \\
                             &                      &                      & \B Mi-2SLpl &  0.024 &  0.012 &  0.041 & [8,155] 160 &  0.008 &  0.010 &  0.042 & [6,141] 145 \\

        \multirow{5}{*}{0.8} & \multirow{5}{*}{0.8} & \multirow{5}{*}{  0} & \T  SimpOLS &  0.584 &  0.343 &  0.040 & - &  0.565 &  0.321 &  0.039 & - \\
                             &                      &                      &      SimpIV &  0.071 &  0.011 &  0.075 & - &  0.056 &  0.009 &  0.074 & - \\
                             &                      &                      &    2SLS-SAR & -0.007 &  0.004 &  0.065 & - & -0.011 &  0.004 &  0.066 & - \\
                             &                      &                      &     Mi-2SLl &  0.090 &  0.022 &  0.041 & [49,144] 158 &  0.056 &  0.015 &  0.041 & [37,138] 148 \\
                             &                      &                      & \B Mi-2SLpl &  0.086 &  0.020 &  0.045 & [49,84] 123 &  0.052 &  0.013 &  0.046 & [37,81] 112 \\

        \multirow{5}{*}{0.8} & \multirow{5}{*}{0.8} & \multirow{5}{*}{0.4} & \T  SimpOLS &  0.586 &  0.345 &  0.039 & - &  0.565 &  0.322 &  0.039 & - \\
                             &                      &                      &      SimpIV &  0.074 &  0.012 &  0.074 & - &  0.058 &  0.009 &  0.073 & - \\
                             &                      &                      &    2SLS-SAR & -0.008 &  0.004 &  0.065 & - & -0.012 &  0.004 &  0.065 & - \\
                             &                      &                      &     Mi-2SLl &  0.100 &  0.023 &  0.043 & [66,125] 152 &  0.064 &  0.016 &  0.042 & [47,127] 144 \\
                             &                      &                      & \B Mi-2SLpl &  0.089 &  0.021 &  0.047 & [66,67] 121 &  0.050 &  0.013 &  0.047 & [47,68] 109 \\
        \hline
     \end{tabular}
     }
  {\footnotesize
  \begin{tablenotes}
      \item \emph{Note}: bias is the bias of $\beta_2$, MSE is the mean squared error of $\beta_2$, AASE is the average asymptotic standard error of $\beta_2$ and [a,b] c is the average number of eigenvectors selected/used in steps 3, 5 and 6 of Algorithm \ref{Mi-2SLalg}.
  \end{tablenotes}
  }
  \end{threeparttable}
\end{table}

\begin{table}[!h]
  \centering
  \caption{Results for $n=500$ and $\omega=0.4$}
  \label{tab:n500,o4}
  \begin{threeparttable}
    {\footnotesize
      \begin{tabular}{ccclrcccrccc}
        \hline
        \multicolumn{3}{c}{Experiment} & \T \B & \multicolumn{4}{c}{Rewiring prob. $p=0.4$} & \multicolumn{4}{c}{Rewiring prob. $p=0.8$} \\
        \T \B $\rho$ & $\zeta_{3,1}$ & $\zeta_{3,2}$ & Estimator & bias & MSE & AASE & Vecs & bias & MSE & AASE  &  Vecs \\
        \hline
        \multirow{5}{*}{0.4} & \multirow{5}{*}{0.4} & \multirow{5}{*}{  0} & \T  SimpOLS &  0.493 &  0.244 &  0.026 & - &  0.482 &  0.233 &  0.025 & - \\
                             &                      &                      &      SimpIV &  0.023 &  0.003 &  0.047 & - &  0.018 &  0.002 &  0.047 & - \\
                             &                      &                      &    2SLS-SAR & -0.009 &  0.002 &  0.046 & - & -0.005 &  0.002 &  0.046 & - \\
                             &                      &                      &     Mi-2SLl &  0.005 &  0.004 &  0.030 & [9,232] 233 &  0.005 &  0.003 &  0.033 & [6,165] 165 \\
                             &                      &                      & \B Mi-2SLpl &  0.010 &  0.004 &  0.031 & [9,192] 200 &  0.008 &  0.003 &  0.034 & [6,138] 143 \\

        \multirow{5}{*}{0.4} & \multirow{5}{*}{0.4} & \multirow{5}{*}{0.4} & \T  SimpOLS &  0.490 &  0.241 &  0.025 & - &  0.479 &  0.230 &  0.025 & - \\
                             &                      &                      &      SimpIV &  0.024 &  0.003 &  0.047 & - &  0.018 &  0.002 &  0.046 & - \\
                             &                      &                      &    2SLS-SAR & -0.009 &  0.002 &  0.045 & - & -0.005 &  0.002 &  0.045 & - \\
                             &                      &                      &     Mi-2SLl &  0.010 &  0.004 &  0.029 & [16,239] 241 &  0.007 &  0.003 &  0.032 & [8,172] 173 \\
                             &                      &                      & \B Mi-2SLpl &  0.015 &  0.004 &  0.031 & [16,180] 193 &  0.009 &  0.003 &  0.034 & [8,138] 145 \\

        \multirow{5}{*}{0.4} & \multirow{5}{*}{0.8} & \multirow{5}{*}{  0} & \T  SimpOLS &  0.503 &  0.253 &  0.024 & - &  0.488 &  0.239 &  0.024 & - \\
                             &                      &                      &      SimpIV &  0.032 &  0.003 &  0.047 & - &  0.024 &  0.003 &  0.046 & - \\
                             &                      &                      &    2SLS-SAR & -0.008 &  0.002 &  0.046 & - & -0.004 &  0.002 &  0.046 & - \\
                             &                      &                      &     Mi-2SLl &  0.018 &  0.005 &  0.035 & [137,74] 189 &  0.023 &  0.005 &  0.033 & [74,113] 164 \\
                             &                      &                      & \B Mi-2SLpl &  0.004 &  0.004 &  0.038 & [137,21] 156 &  0.009 &  0.003 &  0.038 & [74,39] 112 \\

        \multirow{5}{*}{0.4} & \multirow{5}{*}{0.8} & \multirow{5}{*}{0.4} & \T  SimpOLS &  0.499 &  0.250 &  0.024 & - &  0.485 &  0.236 &  0.024 & - \\
                             &                      &                      &      SimpIV &  0.034 &  0.003 &  0.046 & - &  0.025 &  0.003 &  0.046 & - \\
                             &                      &                      &    2SLS-SAR & -0.009 &  0.002 &  0.045 & - & -0.005 &  0.002 &  0.045 & - \\
                             &                      &                      &     Mi-2SLl &  0.009 &  0.004 &  0.036 & [175,43] 203 &  0.020 &  0.004 &  0.034 & [92,92] 162 \\
                             &                      &                      & \B Mi-2SLpl &  0.003 &  0.004 &  0.037 & [175,12] 185 &  0.007 &  0.003 &  0.038 & [92,28] 119 \\

        \multirow{5}{*}{0.8} & \multirow{5}{*}{0.4} & \multirow{5}{*}{  0} & \T  SimpOLS &  0.545 &  0.298 &  0.029 & - &  0.518 &  0.270 &  0.028 & - \\
                             &                      &                      &      SimpIV &  0.051 &  0.005 &  0.052 & - &  0.037 &  0.004 &  0.050 & - \\
                             &                      &                      &    2SLS-SAR & -0.011 &  0.002 &  0.046 & - & -0.007 &  0.002 &  0.046 & - \\
                             &                      &                      &     Mi-2SLl & -0.002 &  0.005 &  0.030 & [9,380] 380 &  0.000 &  0.004 &  0.028 & [6,326] 326 \\
                             &                      &                      & \B Mi-2SLpl &  0.018 &  0.008 &  0.029 & [9,362] 366 &  0.011 &  0.005 &  0.028 & [6,310] 313 \\

        \multirow{5}{*}{0.8} & \multirow{5}{*}{0.4} & \multirow{5}{*}{0.4} & \T  SimpOLS &  0.544 &  0.297 &  0.029 & - &  0.516 &  0.267 &  0.028 & - \\
                             &                      &                      &      SimpIV &  0.052 &  0.006 &  0.052 & - &  0.038 &  0.004 &  0.050 & - \\
                             &                      &                      &    2SLS-SAR & -0.012 &  0.002 &  0.045 & - & -0.007 &  0.002 &  0.045 & - \\
                             &                      &                      &     Mi-2SLl &  0.006 &  0.006 &  0.030 & [16,384] 385 &  0.002 &  0.004 &  0.028 & [8,333] 333 \\
                             &                      &                      & \B Mi-2SLpl &  0.033 &  0.009 &  0.029 & [16,355] 362 &  0.015 &  0.005 &  0.028 & [8,310] 315 \\

        \multirow{5}{*}{0.8} & \multirow{5}{*}{0.8} & \multirow{5}{*}{  0} & \T  SimpOLS &  0.573 &  0.329 &  0.028 & - &  0.537 &  0.290 &  0.027 & - \\
                             &                      &                      &      SimpIV &  0.066 &  0.007 &  0.052 & - &  0.048 &  0.005 &  0.050 & - \\
                             &                      &                      &    2SLS-SAR & -0.011 &  0.002 &  0.046 & - & -0.007 &  0.002 &  0.046 & - \\
                             &                      &                      &     Mi-2SLl &  0.134 &  0.027 &  0.028 & [137,297] 329 &  0.068 &  0.012 &  0.027 & [74,297] 314 \\
                             &                      &                      & \B Mi-2SLpl &  0.120 &  0.023 &  0.031 & [137,153] 259 &  0.065 &  0.011 &  0.031 & [74,161] 226 \\

        \multirow{5}{*}{0.8} & \multirow{5}{*}{0.8} & \multirow{5}{*}{0.4} & \T  SimpOLS &  0.573 &  0.329 &  0.027 & - &  0.536 &  0.288 &  0.026 & - \\
                             &                      &                      &      SimpIV &  0.068 &  0.007 &  0.051 & - &  0.049 &  0.005 &  0.049 & - \\
                             &                      &                      &    2SLS-SAR & -0.011 &  0.002 &  0.046 & - & -0.007 &  0.002 &  0.045 & - \\
                             &                      &                      &     Mi-2SLl &  0.141 &  0.028 &  0.030 & [175,251] 316 &  0.077 &  0.014 &  0.027 & [92,279] 305 \\
                             &                      &                      & \B Mi-2SLpl &  0.122 &  0.024 &  0.032 & [175,128] 265 &  0.064 &  0.011 &  0.032 & [92,135] 217 \\
        \hline
     \end{tabular}
     }
  {\footnotesize
  \begin{tablenotes}
      \item \emph{Note}: bias is the bias of $\beta_2$, MSE is the mean squared error of $\beta_2$, AASE is the average asymptotic standard error of $\beta_2$ and [a,b] c is the average number of eigenvectors selected/used in steps 3, 5 and 6 of Algorithm \ref{Mi-2SLalg}.
  \end{tablenotes}
  }
  \end{threeparttable}
\end{table}

\setlength{\tabcolsep}{6pt} 

\section{Additional application tables }\label{extra_tabs}

\begin{table}[!h]
  \centering
  \caption{Mi-2SL results of \citet{ck16} with 600km SWM cut-off}\
  \label{ckMi-2SLl_600}
  \begin{threeparttable}
    {\footnotesize
    \begin{tabular}{lccccc}
      \hline
      \T  &&&&& Other \\
      \B  & All & Native-born & Foreign-born & Mexican-born & foreign-born \\
      \cline{2-6}
      \multicolumn{6}{l}{\qquad \emph{Panel A: Men, high school or less} \T\B }   \\
  \T  Change in log of group-specific &  0.266 &  0.12 &  0.503 & 1.119 & -0.574 \\
      employment                      & (0.121) & (0.098) & (0.422) & (0.324) & (0.307) \\
      First stage F-statistic         & 62.26 & 44.78 & 74 & 51.03 & 58.52 \\
      Partial F-statistic (Bartik)    & 54.85 & 50.12 & 43.63 & 42.36 & 128.47 \\
  \B  Number of vecs [1st,2nd]        & 9[9,0] & 11[11,0] & 5[5,0] & 14[14,0] & 3[3,0] \\
      \multicolumn{6}{l}{\qquad \emph{Panel B: Men, some college or more} \T\B }   \\
  \T  Change in log of group-specific &  0.431 &  0.556 & -0.256 &  0.92 &  0.025 \\
      employment                      & (0.123) & (0.147) & (0.181) & (0.589) & (0.302) \\
      First stage F-statistic         & 26.96 & 23.74 & 29.76 & 176.09 & 40.02 \\
      Partial F-statistic (Bartik)    & 71.35 & 65.88 & 54.76 & 155.69 & 81.41 \\
  \B  Number of vecs [1st,2nd]        & 9[9,0] & 10[10,0] & 4[4,0] & 15[15,0] & 9[9,0] \\
      \multicolumn{6}{l}{\qquad \emph{Panel C: Women, high school or less} \T\B }   \\
  \T  Change in log of group-specific &  0.151 & -0.22 &  0.272 & 1.811 & -0.656 \\
      employment                      & (0.153) & (0.2) & (0.504) & (0.665) & (0.431) \\
      First stage F-statistic         & 32.88 & 32.52 & 15.21 & 6.04 & 48.69 \\
      Partial F-statistic (Bartik)    & 64.4 & 89.89 & 26.76 & 13.74 & 108.49 \\
  \B  Number of vecs [1st,2nd]        & 1[1,0] & 7[7,0] & 0[0,0] & 0[0,0] & 1[1,0] \\
      \multicolumn{6}{l}{\qquad \emph{Panel D: Women, some college or more} \T\B }   \\
  \T  Change in log of group-specific &  0.039 &  0.011 & -0.357 & -0.054 & -0.405 \\
      employment                      & (0.256) & (0.344) & (0.316) & (1.149) & (0.373) \\
      First stage F-statistic         & 17.75 & 15.12 & 28.86 & 20.02 & 29.07 \\
      Partial F-statistic (Bartik)    & 14.6 & 11.7 & 48.14 & 32.44 & 62.6 \\
  \B  Number of vecs [1st,2nd]        & 1[1,0] & 7[7,0] & 0[0,0] & 0[0,0] & 1[1,0] \\
      \hline
     \end{tabular}
     }
  {\footnotesize
  \begin{tablenotes}
      \item \emph{Note}: First stage fitted values from Lasso estimates (step 3 in Algorithm \ref{Mi-2SLalg}) and the cut-off for the SWM is 600km. Robust standard errors in parentheses.
  \end{tablenotes}
  }
  \end{threeparttable}
\end{table}

\begin{table}[!h]
  \centering
  \caption{Mi-2SL results of \citet{ck16} with 700km SWM cut-off }
  \label{ckMi-2SLl_700}
  \begin{threeparttable}
    {\footnotesize
    \begin{tabular}{lccccc}
      \hline
      \T  &&&&& Other \\
      \B  & All & Native-born & Foreign-born & Mexican-born & foreign-born \\
      \cline{2-6}
      \multicolumn{6}{l}{\qquad \emph{Panel A: Men, high school or less} \T\B }   \\
  \T  Change in log of group-specific &  0.264 &  0.055 &  0.488 & 1.397 & -0.511 \\
      employment                      & (0.133) & (0.093) & (0.361) & (0.287) & (0.302) \\
      First stage F-statistic         & 45.31 & 26.73 & 69.01 & 53.5 & 165.05 \\
      Partial F-statistic (Bartik)    & 82.18 & 83.65 & 40.84 & 51.4 & 157.86 \\
  \B  Number of vecs [1st,2nd]        & 7[7,0] & 14[14,0] & 6[6,0] & 13[13,0] & 4[4,0] \\
      \multicolumn{6}{l}{\qquad \emph{Panel B: Men, some college or more} \T\B }   \\
  \T  Change in log of group-specific &  0.473 &  0.545 & -0.091 & -0.107 &  0.328 \\
      employment                      & (0.15) & (0.136) & (0.21) & (0.747) & (0.252) \\
      First stage F-statistic         & 29.93 & 29.32 & 23.45 & 192.57 & 29.6 \\
      Partial F-statistic (Bartik)    & 88.93 & 138.98 & 68.67 & 119.4 & 126.45 \\
  \B  Number of vecs [1st,2nd]        & 12[12,0] & 13[13,0] & 7[7,0] & 17[17,0] & 10[10,0] \\
      \multicolumn{6}{l}{\qquad \emph{Panel C: Women, high school or less} \T\B }   \\
  \T  Change in log of group-specific &  0.221 &  0.004 &  0.272 & 1.811 & -0.616 \\
      employment                      & (0.129) & (0.235) & (0.504) & (0.665) & (0.367) \\
      First stage F-statistic         & 29.66 & 36.31 & 15.21 & 6.04 & 48.42 \\
      Partial F-statistic (Bartik)    & 72.12 & 60.39 & 26.76 & 13.74 & 140.81 \\
  \B  Number of vecs [1st,2nd]        & 2[2,0] & 8[8,0] & 0[0,0] & 0[0,0] & 2[2,0] \\
      \multicolumn{6}{l}{\qquad \emph{Panel D: Women, some college or more} \T\B }   \\
  \T  Change in log of group-specific &  0.267 &  0.206 & -0.176 & -0.036 & -0.094 \\
      employment                      & (0.218) & (0.288) & (0.327) & (1.134) & (0.342) \\
      First stage F-statistic         & 34.73 & 32.15 & 39.95 & 17.16 & 49.31 \\
      Partial F-statistic (Bartik)    & 30.59 & 27.58 & 53.89 & 31.75 & 66.99 \\
  \B  Number of vecs [1st,2nd]        & 2[2,0] & 8[8,0] & 0[0,0] & 0[0,0] & 2[2,0] \\
      \hline
     \end{tabular}
     }
  {\footnotesize
  \begin{tablenotes}
      \item \emph{Note}: First stage fitted values from Lasso estimates (step 3 in Algorithm \ref{Mi-2SLalg}) and the cut-off for the SWM is 700km. Robust standard errors in parentheses.
  \end{tablenotes}
  }
  \end{threeparttable}
\end{table}

\end{appendices}
\end{document}